\documentclass[showpacs,aps]{revtex4}

\usepackage{amsfonts}
\usepackage{amsmath}
\usepackage{graphicx}

\def\func#1{\mathop{\rm #1}\nolimits}

\def\ii{\'{\i}}

\def\cc{\c{c}}

\begin{document}

\title{Aspects of the competition between atom-field and
field-environment couplings under the influence of an external source in
dispersive
Jaynes-Cummings model}
\author{J. G. Peixoto de Faria \footnote{
Electronic address: jgfaria@fma.if.usp.br} \\
}
\affiliation{
Departamento de F\'{i}sica Matem\'{a}tica, Instituto de F\'{i}sica\\
Universidade de S\~{a}o Paulo\\
C.P. 66318, CEP 05315-970\\
S\~{a}o Paulo, S\~{a}o Paulo, Brazil 
}
\author{
M. C. Nemes
}
\affiliation{
Departamento de F\'{i}sica, Instituto de Ci\^{e}ncias Exatas\\
Universidade Federal de Minas Gerais\\
C.P. 702, CEP\ 30161-970\\
Belo Horizonte, Minas Gerais, Brazil}
\begin{abstract}
We give a fully analytical description of the dynamics of an atom
dispersively coupled to a field mode in a dissipative environment fed by
an external source. The competition between the unitary atom-field (which
leads to entanglement) and the dissipative field-environment couplings are
investigated in detail. We find the time evolution of the global atom-field
system for any intial state and we show that atom-field steady state is at
most classically correlated. For an initial state chosen, we evaluate the
purity loss of the global system and of atomic and field subsystems as
a function of time. We find that the source will tend to compensate for the
dissipation of the field intensity and to accelerate decoherence of
the global and atomic states. Moreover, we show that the degree of
entanglement of the atom-field system, for the particular initial state
chosen, can be completely quantified by concurrence. Analytical
expression for time evolution of the concurrence is given.
\end{abstract}
\maketitle

\section{Introduction}

Quantum mechanics of open systems has lived a revival of interest
especially after the early days of quantum computation \cite{qcomp} when it
has been realized that maintaining quantum coherence is an essential
ingredient to fully exploit the new possibilities opened by the applications
of quantum mechanics in computational physics. Devices using unique quantum
mechanical features can perform information processing in a much more
efficient way than those at work nowadays. The key ingredients of quantum
computing devices with computational capabilities that superseded these
classical counterpart are basically:

\begin{enumerate}
\item  the linear structure of their state space;

\item  the \textit{unitary} character of their dynamical evolution and

\item  the tensorized form of multiparticle states.
\end{enumerate}

The last one, in particular, represents a striking departure from
classicality due to \textit{entanglement}, since combining different systems
results in an exponential growth of the available coding space; moreover,
the tensor product structure is the very basis of many efficient quantum
manipulations.

Of course, all this holds just for closed quantum systems. Real world
quantum systems interact with their environment to a greater or lesser
extent. No matter how weak the coupling to such an environment, the
evolution of quantum subsystems is eventually affected by non-unitary
features such as decoherence, dissipation and heating. From a mathematical
point of view, the relevant state space, given by density matrices, has now
a convex structure and the allowed quantum dynamics is described by
completely positive (CP) maps. Initial pure states preparation are typically
corrupted on extremely short time scales due to quantum coherence loss that
turns them into mixed states: the initial information irreversibly leaks out
from the system into the very large number of uncontrollable degrees of
freedom of the environment.

The above mentioned limitations are of course common to the many devices
proposed for quantum computation \cite{qcomp,devices}. In the present
contribution we will restrict ourselves to a detailed study of cavity
quantum electrodynamics (cavity QED) \cite{cavity-QED}. In this case one
typically has a superconducting cavity with a coherent state fed into it.
The interesting dynamics for quantum computation or for studying
basic features of quantum mechanics is related to the (unitary) atom-field
coupling. It will produce several typical quantum features such as
entanglement and superpositions of states \cite{JCM}. However, it is
impossible to avoid that from the moment a coherent field is fed into the
cavity, it will couple to the environment and the effects of the latter
will tend to
destroy all the typically quantum mechanical features due to the unitary
coupling. One of the aims of the present paper is to investigate the
relation between the time scales of these two competing processes. Another
question which, to our knowledge, has not been satisfactorily answered so
far is the following: the atom-field coupling depends crucially on the
presence of the field in the cavity. If one is working at zero temperature
it is clear that asymptotically the field will go to its vacuum state,
rendering the unitary atom-field entanglement impossible quite independently
of the environment. The basic question is: can one circumvent this problem
by adding an external source which will be present during the whole process?
How does it affect all other time scales of the process?

Besides the fundamental issues addressed by the present work, the results
obtained here can also have purely practical purposes. In fact, recently
de Oliveira, Moussa and Mizrahi \cite{oliveira} proposed a control
mechanism of a mesoscopic superposition of states
(``Schr\"{o}dinger cat" state) in a
dissipative cavity by the coupling with an external source. It is a
variant of the scheme for creation and monitoring of coherent superpositions
of field states described in \cite{cats}. In Ref. \cite{gerry}, Gerry
studied the interaction of an atom with both a quantized cavity field and an
external classical field. The interaction between atom and cavity
field has a dispersive character. As a result, various forms of superpositions
of cavity field states can be produced. Further, robust coherent states
may be generated in the steady state of the cavity if dissipation
is included. The present work adds to those contributions in the sense
that the source is treated quantum mechanically and dissipation is taken
into account while atom and field interact. We show that the presence of an
external source will attenuate dissipative effects, since it will maintain
a constant field intensity in the cavity. However, since decoherence
strongly depends on the field intensity, the presence of an external
source will tend to increase typical decoherence time scales.

Moreover, a rather new aspect treated here is the quantification of
quantum correlations in this dissipative dynamics. It is not difficult to
find measures of the degree of entanglement between subsystems of a
global system whose state is pure. The
situation dramatically changes, however, if the global state
is characterized by a statistical mixture. For two qubits systems it
is possible to provide for a quantitative measure of entanglement using
the notion of concurrence \cite{concurrence}. For a particular
initial state, we show that the atom-field system can be mapped onto
a system composed by two two-dimensional subsystems. We therefore give
analytical expressions for the time evolution of the entanglement
between atom and field measured by the concurrence. To our knowledge,
it is the first time that the time evolution of the concurrence is
obtained for a dissipative system.

This article was organized as follows: in Section II we find the time
development of the atom-field state for any initial condition. In
Section III, we calculate the time evolution of an initial state chosen and
we obtain the idempotency defect or linear entropy of global
and reduced density operators
in order to study the purity loss of the compound system and of the
atomic and field subsystems. We obtain correlation measures of
the global state. In particular, the degree of entanglement between
atom and field is evaluated by finding an expression for concurrence.
Finally, the appendix describes the method used to obtain
expressions for the solutions of different equations of motion
that appear in Section II.

\section{The dynamics of the system atom-field in the presence of a resonant
external source}

Before handling the complete problem of the dispersive atom-field
interaction in a dissipative environment with an external source, let us
consider the time evolution of a single driven electromagnetic field mode
coupled to a thermal reservoir at null temperature. The evolution of the
field state, described by the density operator $\hat{\rho }_{f}$ in the
interaction picture, is governed by the master equation

\begin{align}
\frac{d}{dt}\hat{\rho }_{f}\left( t\right) &= -i\left[ Fe^{-i\left(
\omega _{S}-\omega _{0}\right) t}\hat{a}^{\dagger }+F^{\ast }e^{i\left(
\omega _{S}-\omega _{0}\right) t}\hat{a},\hat{\rho }_{f}\left(
t\right) \right]  \label{master1} \\
& +k\left[ 2\hat{a}\hat{\rho }_{f}\left( t\right) \hat{a}%
^{\dagger }-\hat{a}^{\dagger }\hat{a}\hat{\rho }_{f}\left(
t\right) -\hat{\rho }_{f}\left( t\right) \hat{a}^{\dagger }\hat{a%
}\right] ,  \notag
\end{align}
where $k$ is the damping constant and $\omega _{0}$ and $\omega _{S}$ are
the mode and source frequencies, respectively. The contribution of the
external source is identified by the commutator in the right hand side of
Eq. (\ref{master1}); the coupling between source and field is given by the
constant $F$. Without loss of generality, let us assume that the source is
resonant with the mode, i.e., $\omega _{S}=\omega _{0}$. Hence, the master
equation can be written as

\begin{equation}
\frac{d}{dt}\hat{\rho }_{f}\left( t\right) =\left( \mathcal{L}_{S}+%
\mathcal{D}\right) \hat{\rho }_{f}\left( t\right) ,  \label{master2}
\end{equation}
where we have defined 
\begin{equation}
\mathcal{L}_{S}\cdot \equiv -i\left[ F\hat{a}^{\dagger }+F^{\ast }%
\hat{a},\cdot \right] =-iF\left[ \left( \hat{a}^{\dagger }\cdot
\right) -\left( \cdot \hat{a}^{\dagger }\right) \right] -iF^{\ast }\left[
\left( \hat{a}\cdot \right) -\left( \cdot \hat{a}\right) \right]
\label{liouvillianS}
\end{equation}
and 
\begin{equation}
\mathcal{D}\equiv k\left( 2\mathcal{J}-\mathcal{M}-\mathcal{P}\right) .
\label{dissipator}
\end{equation}
The superoperators in Eq. (\ref{dissipator}) are $\mathcal{M}\cdot \equiv 
\hat{a}^{\dagger }\hat{a}\cdot $, $\mathcal{P}\cdot \equiv \cdot 
\hat{a}^{\dagger }\hat{a}$, $\mathcal{J}\cdot \equiv \hat{a}%
\cdot \hat{a}^{\dagger }$. The term $\mathcal{D}$ is called
``dissipator'' and contains the non-unitary contributions to the dynamics of 
$\hat{\rho }_{f}$. The formal solution of the equation (\ref{master1})
is given by the expression 
\begin{equation}
\hat{\rho }_{f}\left( t\right) =e^{\left( \mathcal{L}_{S}+\mathcal{D}%
\right) t}\hat{\rho }_{f}\left( 0\right) ,  \label{formal1}
\end{equation}
where $\hat{\rho }_{f}\left( 0\right) $ represents the field state at $%
t=0$. The solution (\ref{formal1}) can be rewritten as 
\begin{equation}
\hat{\rho }_{f}\left( t\right) =\hat{D}\left[ -i\frac{F}{k}\left(
1-e^{-kt}\right) \right] \left[ \exp \left( \mathcal{D}t\right) \hat{%
\rho }_{f}\left( 0\right) \right] \hat{D}^{\dagger }\left[ -i\frac{F}{k}%
\left( 1-e^{-kt}\right) \right] .  \label{rhoF2}
\end{equation}
Here, $\hat{D}$ is the displacement operator of the Heisenberg-Weyl
group, defined as \cite{perelomov} 
\begin{equation}
\hat{D}\left( \alpha \right) =\exp \left( \alpha \hat{a}^{\dagger
}-\alpha ^{\ast }\hat{a}\right) =\exp \left( -\frac{\left| \alpha
\right| ^{2}}{2}\right) e^{\alpha \hat{a}^{\dagger }}e^{-\alpha ^{\ast }%
\hat{a}}=\exp \left( \frac{\left| \alpha \right| ^{2}}{2}\right)
e^{-\alpha ^{\ast }\hat{a}}e^{\alpha \hat{a}^{\dagger }}.
\label{displacement}
\end{equation}

Note that the coherent state 
\begin{equation}
\hat{\rho }_{f}^{\mathrm{stat}}=\left| -iF/k\right\rangle \left\langle
-iF/k\right|  \label{stat1}
\end{equation}
is the stationary state of the dynamics described by the master equation (%
\ref{master2}). In fact, it is easy to show that $\left( \mathcal{L}_{S}+%
\mathcal{D}\right) \hat{\rho }_{f}^{\mathrm{stat}}=0$. Besides, since $%
\lim_{t\rightarrow \infty }\exp \left( \mathcal{D}t\right) \hat{\rho }%
_{f}\left( 0\right) =\left| 0\right\rangle \left\langle 0\right| $,
whichever it may be the initial state $\hat{\rho }_{f}\left( 0\right) $,
then $\hat{\rho }_{f}\left( t\rightarrow \infty \right) =\hat{\rho }%
_{f}^{\mathrm{stat}}$. In the other words, in the limit $t\rightarrow \infty 
$, the field state converges to the state $\hat{\rho }_{f}^{\mathrm{stat}%
}$. Therefore, as a result of the coupling between the field mode and an
external source, a stationary coherent state is produced.

\subsection{Solution of the equations of motion}

From now on, we will consider the field interacting dispersively with a two
level atom and coupled to a zero temperature reservoir and to an external
source. In the interaction picture, the evolution of the compound 
atom-field system, described by the density operator $\hat{\rho }$, is governed
by the master equation 
\begin{align}
\frac{d}{dt}\hat{\rho }\left( t\right) &= -i\omega \left[ \left( 
\hat{a}^{\dagger }\hat{a}+1\right) \left| e\right\rangle
\left\langle e\right| -\hat{a}^{\dagger }\hat{a}\left|
g\right\rangle \left\langle g\right| ,\hat{\rho }\left( t\right) \right]
\label{master3} \\
& -i\left[ F\hat{a}^{\dagger }+F^{\ast }\hat{a},\hat{\rho }%
\left( t\right) \right]  \notag \\
& +k\left[ 2\hat{a}\hat{\rho }\left( t\right) \hat{a}^{\dagger }-%
\hat{a}^{\dagger }\hat{a}\hat{\rho }\left( t\right) -\hat{%
\rho }\left( t\right) \hat{a}^{\dagger }\hat{a}\right] .  \notag
\end{align}
$e$ and $g$ are the atomic levels of interest and $\omega \equiv
G^{2}/\delta $, where $G$ measures the coupling between atom and field and $%
\delta $ is the difference between the frequency of the atomic transition
and the frequency of the mode (detuning). In order that the dispersive
approximation remains valid in the presence of the source, the condition $%
\left| \delta \right| /G\gg \left| F\right| /k$ must be satisfied.

Equations of motion for the operators $\hat{\rho }_{ee}\left( t\right)
=\left\langle e\right| \hat{\rho }\left( t\right) \left| e\right\rangle $%
, $\hat{\rho }_{gg}\left( t\right) =\left\langle g\right| \hat{\rho }%
\left( t\right) \left| g\right\rangle $ and $\hat{\rho }_{ge}\left(
t\right) =\hat{\rho }_{eg}^{\dagger }\left( t\right) =\left\langle
g\right| \hat{\rho }\left( t\right) \left| e\right\rangle $ are obtained
from the master equation (\ref{master3}). These equations have the general
form 
\begin{equation}
\frac{d}{dt}\hat{\rho }_{ij}\left( t\right) =\mathcal{L}_{ij}^{\prime }%
\hat{\rho }_{ij}\left( t\right) ,  \label{gen_form}
\end{equation}
where $\hat{\rho }_{ij}=\left\langle i\right| \hat{\rho }\left(
t\right) \left| j\right\rangle $, $i,j=\left( e,g\right) $, represents any
of the operators above defined, and $\mathcal{L}_{ij}^{\prime }$ is the
respective dynamical generator. $\mathcal{L}_{ij}^{\prime }$ can be written
as 
\begin{equation}
\mathcal{L}_{ij}^{\prime }=\mathcal{L}_{ij}+\mathcal{L}_{S},
\label{liouvillian_ij}
\end{equation}
where $\mathcal{L}_{ij}$ represents one of the superoperators 
\begin{align}
\mathcal{L}_{gg} &\equiv \mathcal{L}_{-}=i\omega \left( \mathcal{M}-%
\mathcal{P}\right) +k\left( 2\mathcal{J}-\mathcal{M}-\mathcal{P}\right) , 
\notag \\
\mathcal{L}_{ee} &\equiv \mathcal{L}_{+}=-i\omega \left( \mathcal{M}-%
\mathcal{P}\right) +k\left( 2\mathcal{J}-\mathcal{M}-\mathcal{P}\right) ,
\label{liouvillians} \\
\mathcal{L}_{eg} &= -i\omega \left( \mathcal{M}+\mathcal{P}+1\right)
+k\left( 2\mathcal{J}-\mathcal{M}-\mathcal{P}\right) ,  \notag
\end{align}
and $\mathcal{L}_{S}$ was defined in (\ref{liouvillianS}).

The formal solution of (\ref{gen_form}) is 
\begin{equation}
\hat{\rho }_{ij}\left( t\right) =e^{\mathcal{L}_{ij}^{\prime }t}\hat{%
\rho }_{ij}\left( 0\right) ,  \label{formal2}
\end{equation}
where $\hat{\rho }_{ij}\left( 0\right) $ represents the operator $%
\hat{\rho }_{ij}$ at $t=0$. In the appendix we describe in details the
method employed to obtain the expressions for the ``matrix elements'' of $%
\hat{\rho }$ that we will show next. The ``diagonal elements'', $%
\hat{\rho }_{ee}\left( t\right) $ and $\hat{\rho }_{gg}\left(
t\right) $, can be expressed as 
\begin{subequations}
\label{diagonal1}
\begin{align}
\hat{\rho }_{ee}\left( t\right) &=\hat{D}\left[ \beta _{e}\left(
t\right) \right] \left[ \exp \left( \mathcal{L}_{+}t\right) \hat{\rho }%
_{ee}\left( 0\right) \right] \hat{D}^{\dagger }\left[ \beta _{e}\left(
t\right) \right] ,  \label{rhoee1} \\
\hat{\rho }_{gg}\left( t\right) &=\hat{D}\left[ \beta _{g}\left(
t\right) \right] \left[ \exp \left( \mathcal{L}_{-}t\right) \hat{\rho }%
_{gg}\left( 0\right) \right] \hat{D}^{\dagger }\left[ \beta _{g}\left(
t\right) \right] .  \label{rhogg1}
\end{align}
\end{subequations}
The arguments of the above displacement operators are 
\begin{subequations}
\label{betas}
\begin{align}
\beta _{e}\left( t\right) &\equiv \frac{F}{\omega -ik}\left[ e^{-\left(
k+i\omega \right) t}-1\right] ,  \label{betae} \\
\beta _{g}\left( t\right) &\equiv -\frac{F}{\omega +ik}\left[ e^{-\left(
k-i\omega \right) t}-1\right] .  \label{betag}
\end{align}
\end{subequations}
The expressions (\ref{rhoee1}) and (\ref{rhogg1}) are completely general.
For a given initial state, the technique presented in the appendix can be
used to determine the operators $\exp \left( \mathcal{L}_{+}t\right) 
\hat{\rho }_{ee}\left( 0\right) $ and $\exp \left( \mathcal{L}%
_{-}t\right) \hat{\rho }_{gg}\left( 0\right) $.

The general solution of the equation of motion for the ``non-diagonal
element'' $\hat{\rho }_{eg}=\hat{\rho }_{ge}^{\dagger }$ can be
written as 
\begin{align}
\hat{\rho }_{eg}\left( t\right) &= \exp \left[ z\left( t\right) +\left|
F\right| ^{2}\left( p^{2}-q^{2}+2pq+\left| p+q\right| ^{2}\right) \left(
t\right) \right] \hat{D}\left[ \beta _{e}\left( t\right) \right]  \notag
\\
& \times \exp \left\{ 2F^{\ast }\left[ \mathop{\rm Re}\nolimits p\left(
t\right) -i\mathop{\rm Im}\nolimits q\left( t\right) \right] \hat{a}%
\right\} \left[ \exp \left( \mathcal{L}_{eg}t\right) \hat{\rho }%
_{eg}\left( 0\right) \right]  \label{rhoeg1} \\
& \times \exp \left\{ -2F\left[ \mathop{\rm Re}\nolimits p\left( t\right) -i%
\mathop{\rm Im}\nolimits q\left( t\right) \right] \hat{a}^{\dagger
}\right\} \hat{D}^{\dagger }\left[ \beta _{g}\left( t\right) \right] . 
\notag
\end{align}
The functions $z\left( t\right) $, $p\left( t\right) $ and $q\left( t\right) 
$ in the above expression are given by 
\begin{subequations}
\label{functions}
\begin{align}
z\left( t\right) &= -\frac{2i\omega \left| F\right| ^{2}}{\left( k+i\omega
\right) ^{2}}\left\{ t+\frac{4\left[ e^{-\left( k+i\omega \right) t}-1\right]
-e^{-2\left( k+i\omega \right) t}+1}{2\left( k+i\omega \right) }\right.
\label{z} \\
& \left. +i\frac{\omega }{\left( k+i\omega \right) ^{2}}\left\{ \cosh \left[
\left( k+i\omega \right) t\right] -1\right\} ^{2}\right\} ,  \notag \\
p\left( t\right) &= i\frac{k}{\left( k+i\omega \right) ^{2}}\left\{ \cosh %
\left[ \left( k+i\omega \right) t\right] -1\right\} -i\frac{\sinh \left[
\left( k+i\omega \right) t\right] }{k+i\omega },  \label{p} \\
q\left( t\right) &= -\frac{\omega }{\left( k+i\omega \right) ^{2}}\left\{
\cosh \left[ \left( k+i\omega \right) t\right] -1\right\} .  \label{q}
\end{align}
\end{subequations}

The analysis of the argument of the first exponential in RHS of Eq. (\ref
{rhoeg1}) provides some understanding of the asymptotic behavior of the
term $\hat{\rho }_{eg}$. The functions defined in (\ref{z})-(\ref{q})
produce 
\begin{align}
z\left( t\right) +\left| F\right| ^{2}\left( \cdots \right) 
\left( t\right) &=-%
\frac{2i\omega \left| F\right| ^{2}}{\left( k+i\omega \right) ^{2}}\left\{ t+%
\frac{4\left[ e^{-\left( k+i\omega \right) t}-1\right] -e^{-2\left(
k+i\omega \right) t}+1}{2\left( k+i\omega \right) }\right\}  \notag \\
& -\frac{\left| F\right| ^{2}}{\left( k+i\omega \right) ^{2}}\left[
e^{-\left( k+i\omega \right) t}-1\right] ^{2}  \label{arg} \\
& +\frac{\left| F\right| ^{2}}{k^{2}+\omega ^{2}}\left(
e^{-2kt}-2e^{-kt}\cos \omega t+1\right) ,  \notag
\end{align}
where $\cdots =p^{2}-q^{2}+2pq+\left| p+q\right| ^{2}$. For long times ($t\gg
1/k $), the RHS of (\ref{arg}) depends linearly on $t$. The real part of the
dominant term in this regime is $-4\omega ^{2}k\left| F\right| ^{2}t/\left(
k^{2}+\omega ^{2}\right) ^{2}$. It is responsible by the complete
disappearance of the ``non-diagonal elements'' $\hat{\rho }_{eg}$ and $%
\hat{\rho }_{ge}$. Therefore, whichever may be the initial state of the
compound system atom-field, the stationary regime is characterized by the
state 
\begin{equation}
\hat{\rho }^{\mathrm{stat}}=\func{tr}_{f} \left[
\hat{\rho }_{ee}\left( 0 \right)\right] \left|F/\left(ik - \omega \right)
\right\rangle \left\langle F/\left(ik - \omega \right) \right|
\otimes \left| e\right\rangle \left\langle e\right| +
\func{tr}_{f}\left[\hat{\rho }
_{gg}\left( 0\right)\right] \left|F/\left(ik + \omega \right)
\right\rangle \left\langle F/\left(ik + \omega \right) \right|
\otimes \left| g\right\rangle \left\langle
g\right| .  \label{stat2}
\end{equation}

The characteristic time of decay of the ``non-diagonal elements'' for long
times depends on the damping constant $k$, the effective coupling between
atom and field $\omega $ and the intensity of the external source $F$. It is
easy to verify that the larger the value of $\left| F\right| $ the faster
the decay of $\hat{\rho }_{eg}$ and $\hat{\rho }_{ge}$. On the other
hand, keeping the value of intensity constant, the decay of the
``non-diagonal elements'' is more rapid in the critical damping ($k/\omega
=1 $) regime for long times.

Finally, in absence of the external source, i.e., making $F=0$ in Eqs. (\ref
{rhoee1}), (\ref{rhogg1}) and (\ref{rhoeg1}), we recover the results
obtained in Ref. \cite{faria}.

\section{The evolution of an uncorrelated initial state}

In order to understand the influence on the entanglement process between
atom and field in dispersive JCM by the introduction of both dissipative
mechanism and external source, let us calculate the time evolution of an
uncorrelated initial state 
\begin{equation}
\left| \psi \right\rangle =\frac{1}{\sqrt{2}}\left( \left| e\right\rangle
+\left| g\right\rangle \right) \otimes \left| -iF/k\right\rangle .
\label{initial_state}
\end{equation}
This state is prepared turning on the source at $t\rightarrow -\infty $.The
source is maintained continuously pumping the field and, at $t=0$, the field
reaches the stationary state (\ref{stat1}). The atom, prepared in a coherent
superposition of the states $\left| e\right\rangle $ and $\left|
g\right\rangle $, begins to interact with the field. Such state was chosen
by simplicity but the results obtained can be easily extended to more general
states such as $\left( \sin \phi \left| e\right\rangle +e^{i\chi }\cos \phi
\left| g\right\rangle \right) \otimes \left| \alpha \right\rangle $.

\subsection{The global density operator}

Given the initial state (\ref{initial_state}), the ``matrix elements'' of $%
\hat{\rho }\left( 0\right) $ are $\hat{\rho }_{ee}\left( 0\right) =%
\hat{\rho }_{gg}\left( 0\right) =\hat{\rho }_{eg}\left( 0\right) =%
\hat{\rho }_{ge}\left( 0\right) =\frac{1}{2}\left| -iF/k\right\rangle
\left\langle -iF/k\right| $. The state $\hat{\rho }\left( t\right) $ is
determined after finding the time evolution of each ``matrix elements''. The
evolved global state is given by the expression 
\begin{equation}
\hat{\rho }\left( t\right) =\hat{\rho }_{gg}\left( t\right) \otimes
\left| g\right\rangle \left\langle g\right| +\hat{\rho }_{ge}\left(
t\right) \otimes \left| g\right\rangle \left\langle e\right| +\hat{\rho }%
_{eg}\left( t\right) \otimes \left| e\right\rangle \left\langle g\right| +%
\hat{\rho }_{ee}\left( t\right) \otimes \left| e\right\rangle
\left\langle e\right| .  \label{rhoAF1}
\end{equation}
Equations (\ref{rhoee1}) and (\ref{rhogg1}) produce, for the ``diagonal
elements'', 
\begin{subequations}
\label{diagonal2}
\begin{align}
\hat{\rho }_{ee}\left( t\right) &= \frac{1}{2}\left| \beta _{e}^{\prime
}\left( t\right) \right\rangle \left\langle \beta _{e}^{\prime }\left(
t\right) \right| ,  \label{rhoee2_fonte} \\
\hat{\rho }_{gg}\left( t\right) &= \frac{1}{2}\left| \beta _{g}^{\prime
}\left( t\right) \right\rangle \left\langle \beta _{g}^{\prime }\left(
t\right) \right| .  \label{rhogg2_fonte}
\end{align}
\end{subequations}
where we have defined 
\begin{subequations}
\label{betas_prime}
\begin{align}
\beta _{e}^{\prime }\left( t\right) &= \beta _{e}\left( t\right) -i\frac{F}{k%
}e^{-\left( k+i\omega \right) t},  \label{betae_prime} \\
\beta _{g}^{\prime }\left( t\right) &= \beta _{g}\left( t\right) -i\frac{F}{k%
}e^{-\left( k-i\omega \right) t}.  \label{betag_prime}
\end{align}
\end{subequations}
Here, $\beta _{e}\left( t\right) $ and $\beta _{g}\left( t\right) $ are the
amplitudes defined in Eqs. (\ref{betae}) and (\ref{betag}).

The evolution of the ``non-diagonal elements'' is calculated by the using of
Eq. (\ref{rhoeg1}): 
\begin{equation}
\hat{\rho }_{eg}\left( t\right) =\frac{1}{2}\exp \left[ \Phi \left(
\omega ,k,F;t\right) \right] \left| \beta _{e}^{\prime }\left( t\right)
\right\rangle \left\langle \beta _{g}^{\prime }\left( t\right) \right| .
\label{rhoeg2}
\end{equation}
The complex phase $\Phi \left( \omega ,k,F;t\right) $ is given by the
expression 
\begin{align}
\Phi \left( \omega ,k,F;t\right) &= -i\omega t+z\left( t\right) +\left|
F\right| ^{2}\left( p^{2}-q^{2}+2pq+\left| p+q\right| ^{2}\right) \left(
t\right)  \label{phi} \\
& +i\Theta \left( F/k,t\right) +\Gamma \left( F/k,t\right)  \notag \\
& +\frac{\left| F\right| ^{2}}{k}\left\{ 2i\mathop{\rm Re}\nolimits\left[
\left( p+q\right) \left( t\right) e^{-kt}\cos \omega t\right] 
-2i\func{Im} \left[\left(p + q\right)\left(t \right) e^{-kt}
\sin \omega t\right]\right.  \notag \\
& \left. -4e^{-\left( k+i\omega \right) t}\left[ \mathop{\rm Im}\nolimits
q\left( t\right) +i\mathop{\rm Re}\nolimits p\left( t\right) \right]
\right\} ,  \notag
\end{align}
where the functions $\Theta \left( F/k,t\right) $ and $\Gamma \left(
F/k,t\right) $ are 
\begin{align*}
\Theta \left( F/k,t\right) &= \frac{\left| F\right| ^{2}}{k\left(
k^{2}+\omega ^{2}\right) }\left[ e^{-2kt}\left( k\sin 2\omega t+\omega \cos
2\omega t\right) -\omega \right] , \\
\Gamma \left( F/k,t\right) &= -\frac{\left| F\right| ^{2}}{k^{2}}\left(
1-e^{-2kt}\right) -\frac{\left| F\right| ^{2}}{k\left( k^{2}+\omega
^{2}\right) } \\
& \times \left[ e^{-2kt}\left( k\cos 2\omega t-\omega \sin 2\omega t\right)
-k\right] .
\end{align*}
Hence, after $t$, the state of the compound system atom-field is 
\begin{align}
\hat{\rho }\left( t\right) &= \frac{1}{2}\left\{ \left| e,\beta
_{e}^{\prime }\left( t\right) \right\rangle \left\langle e,\beta
_{e}^{\prime }\left( t\right) \right| +\left| g,\beta _{g}^{\prime }\left(
t\right) \right\rangle \left\langle g,\beta _{g}^{\prime }\left( t\right)
\right| \right.  \label{rhoAF2} \\
& \left. +\left\{ \exp \left[ \Phi \left( \omega ,k,F;t\right) \right]
\left| e,\beta _{e}^{\prime }\left( t\right) \right\rangle \left\langle
g,\beta _{g}^{\prime }\left( t\right) \right| +\mathrm{H.\ c.}\right\}
\right\} ,  \notag
\end{align}
where $\mathrm{H.\ c.}$ stands for Hermitean conjugate.

The density operator $\hat{\rho }\left( t\right) $ can be diagonalized
yielding the following eigenvectors and eigenvalues
\begin{subequations}
\label{EV_global} 
\begin{align}
\left| \psi _{+}\right\rangle &= \frac{1}{\sqrt{2}}\left\{ \left| e,\beta
_{e}^{\prime }\left( t\right) \right\rangle +\exp \left[ -i\mathop{\rm Im}%
\nolimits\Phi \left( \omega ,k,F;t\right) \right] \left| g,\beta
_{g}^{\prime }\left( t\right) \right\rangle \right\} ,
\label{EV1_global} \\
\left| \psi _{-}\right\rangle &= \frac{1}{\sqrt{2}}\left\{ \left| e,\beta
_{e}^{\prime }\left( t\right) \right\rangle -\exp \left[ -i\mathop{\rm Im}%
\nolimits\Phi \left( \omega ,k,F;t\right) \right] \left| g,\beta
_{g}^{\prime }\left( t\right) \right\rangle \right\} ,
\label{EV2_global}
\end{align}
\end{subequations}
\begin{subequations}
\label{ev_global}
\begin{align}
\lambda _{+}\left( t\right) &= \frac{1}{2}\left\{ 1+\exp \left[ \mathop{\rm
Re}\nolimits\Phi \left( \omega ,k,F;t\right) \right] \right\} ,
\label{ev1_global} \\
\lambda _{-}\left( t\right) &= \frac{1}{2}\left\{ 1-\exp \left[ \mathop{\rm
Re}\nolimits\Phi \left( \omega ,k,F;t\right) \right] \right\} ,
\label{ev2_global}
\end{align}
\end{subequations}
in terms of which $\hat{\rho }\left( t\right) $ can be written as 
\begin{equation*}
\hat{\rho }\left( t\right) =\lambda _{+}\left( t\right) \left| \psi
_{+}\right\rangle \left\langle \psi _{+}\right| +\lambda _{-}\left( t\right)
\left| \psi _{-}\right\rangle \left\langle \psi _{-}\right| .
\end{equation*}

The purity of the state represented by a density operator $\hat{\rho }$
is conveniently measured by the idempotency defect or linear entropy \cite
{lin-ent} 
\begin{equation*}
\varsigma \left( t\right) =1-\mathop{\rm tr}\hat{\rho }^{2}\left(
t\right)
\end{equation*}
In general, if $\hat{\rho }$ describes a pure state, $\varsigma =0$,
otherwise $\varsigma >0$. The idempotency defect of the state of the system
atom-field as a function of time is given by 
\begin{equation}
\varsigma \left( t\right) =1-\lambda _{+}^{2}\left( t\right) -\lambda
_{-}^{2}\left( t\right) =\frac{1}{2}\left\{ 1-\exp \left[ 2\mathop{\rm Re}%
\nolimits\Phi \left( \omega ,k,F;t\right) \right] \right\} .
\label{linear_ent_global}
\end{equation}
The idempotency defect has an upper limit which is characteristic of a
complete mixture. In the case studied, this value is $1/2$. As we already
discussed, the ``non-diagonal elements'' $\hat{\rho }_{eg}\left(
t\right) =\hat{\rho }_{ge}^{\dagger }\left( t\right) $ vanish at $%
t\rightarrow \infty $. As consequence, the coherence loss of the system
atom-field is complete in the stationary regime, i.e., $\varsigma \left(
t\rightarrow \infty \right) =1/2.$

\subsubsection{Short and long times behavior of coherence loss of global
state}

The real part of the function $\Phi \left( \omega ,k,F;t\right) $ controls
the behavior of the linear entropy $\varsigma $ of the state of the global
system $\hat{\rho }$. For long times ($kt\gg 1$), $\Phi $ depends
linearly on $t$, and $\varsigma $ grows accordingly $\varsigma \left(
t\right) \sim 1-\exp \left( -2t/\tau _{\mathrm{dec}}^{\mathrm{lt}}\right) $.
The characteristic decoherence time in this regime is 
\begin{equation}
\tau _{\mathrm{dec}}^{\mathrm{lt}}=\frac{\left( k^{2}+\omega ^{2}\right) ^{2}%
}{4\omega ^{2}k\left| F\right| ^{2}}=\left\{ kD^{2}\left[ \beta _{e}^{\prime
}\left( \infty \right) ,\beta _{g}^{\prime }\left( \infty \right) \right]
\right\} ^{-1},  \label{lt_char}
\end{equation}
where $D\left[ \beta _{e}^{\prime }\left( \infty \right) ,\beta _{g}^{\prime
}\left( \infty \right) \right] $ stands for the distance in the phase space
between the coherent states $\left| \beta _{e}^{\prime }\left( \infty
\right) \right\rangle $ and $\left| \beta _{g}^{\prime }\left( \infty
\right) \right\rangle $. This distance , defined by the expression $D\left(
z,z^{\prime }\right) =\left| z-z^{\prime }\right| $, measures the
distinguishability between the coherent states $\left| z\right\rangle $ and $%
\left| z^{\prime }\right\rangle $ ($z$ and $z^{\prime }$ are two complex
numbers). The amplitudes $\beta _{e}^{\prime }\left( \infty \right) $ and $%
\beta _{g}^{\prime }\left( \infty \right) $ are given by 
\begin{equation}
\beta _{e}^{\prime }\left( \infty \right) =-\frac{F}{\omega -ik},\qquad
\beta _{g}^{\prime }\left( \infty \right) =\frac{F}{\omega +ik}.
\label{betas_inf}
\end{equation}
Hence, for long times, the more distinguishable the stationary states $%
\left| \beta _{e}^{\prime }\left( \infty \right) \right\rangle $ and $\left|
\beta _{g}^{\prime }\left( \infty \right) \right\rangle $, the more rapid
the decoherence process of the global state. This result is a direct
consequence of the long times behavior of the decay of the ``non-diagonal
elements'' discussed above. In fact, the characteristic decoherence time of
the global state inherits the properties of the characteristic time of decay
of the ``non-diagonal elements'' for long times.

For short times ($kt\ll 1$ and $\omega t\ll 1$), the linear entropy $%
\varsigma $ grows accordingly $\varsigma \left( t\right) \sim 1-\exp \left[
-2\left( t/\tau _{\mathrm{dec}}^{\mathrm{st}}\right) ^{3}\right] $, where 
\begin{equation}
\left( \tau _{\mathrm{dec}}^{\mathrm{st}}\right) ^{-3}=\left| F\right| ^{2}k%
\left[ 1+\frac{4}{3}\left( \frac{\omega }{k}\right) ^{2}\right] .
\label{st_char}
\end{equation}
In this regime, the larger the intensity of the external source $\left|
F\right| ^{2}$, the faster is the coherence loss. Besides, for a given
intensity value, the global state begins to lose coherence more rapidly in
the subcritical damping regime ($k/\omega <1$).

\subsection{The reduced density operators}

The state of the atomic subsystem (respectively, field subsystem) is
described by the density operator $\hat{\rho }_{a}$ (respectively, $%
\hat{\rho }_{f}$). This operator is obtained by taking the partial trace of 
$\hat{\rho }$ with respect to the field variables
(respectively, atomic variables).
The atomic density operator is given by 
\begin{align}
\hat{\rho }_{a}\left( t\right) &= \mathop{\rm tr}\nolimits_{f}\hat{%
\rho }\left( t\right)  \notag \\
&= \frac{1}{2}\left\{ \left| e\right\rangle \left\langle e\right| +\left|
g\right\rangle \left\langle g\right| \right.  \label{rhoA2} \\
& \left. +\left\{ \exp \left[ \Phi \left( \omega ,k,F;t\right) \right]
\left\langle \beta _{g}^{\prime }\left( t\right) \right| \left. \beta
_{e}^{\prime }\left( t\right) \right\rangle \left| e\right\rangle
\left\langle g\right| +\mathrm{H.\ c.}\right\} \right\} .  \notag
\end{align}
This operator can be diagonalized, yielding the following eigenvectors and
eigenvalues 
\begin{subequations}
\label{EV_atom}
\begin{align}
\left| g^{\prime }\right\rangle &= \frac{1}{\sqrt{2}}\left\{ \left|
e\right\rangle +\exp \left\{ -i\mathop{\rm Im}\nolimits\left[ \Phi \left(
\omega ,k,F;t\right) +\beta _{e}^{\prime }\left( t\right) \beta _{g}^{\prime
\ast }\left( t\right) \right] \right\} \left| g\right\rangle \right\} ,
\label{EV1_atom} \\
\left| e^{\prime }\right\rangle &= \frac{1}{\sqrt{2}}\left\{ \left|
e\right\rangle -\exp \left\{ -i\mathop{\rm Im}\nolimits\left[ \Phi \left(
\omega ,k,F;t\right) +\beta _{e}^{\prime }\left( t\right) \beta _{g}^{\prime
\ast }\left( t\right) \right] \right\} \left| g\right\rangle \right\} ,
\label{EV2_atom}
\end{align}
\end{subequations}
\begin{subequations}
\label{ev_atom}
\begin{align}
\lambda _{g^{\prime }}\left( t\right) &= \frac{1}{2}\left\{ 1+\exp \left\{ %
\mathop{\rm Re}\nolimits\Phi \left( \omega ,k,F;t\right) -\frac{1}{2}D^{2}%
\left[ \beta _{e}^{\prime }\left( t\right) ,\beta _{g}^{\prime }\left(
t\right) \right] \right\} \right\} ,  \label{ev1_atom} \\
\lambda _{e^{\prime }}\left( t\right) &= \frac{1}{2}\left\{ 1-\exp \left\{ %
\mathop{\rm Re}\nolimits\Phi \left( \omega ,k,F;t\right) -\frac{1}{2}D^{2}%
\left[ \beta _{e}^{\prime }\left( t\right) ,\beta _{g}^{\prime }\left(
t\right) \right] \right\} \right\} .  \label{ev2_atom}
\end{align}
\end{subequations}
The atomic purity loss is measured by the idempotency defect: 
\begin{equation}
\varsigma _{a}\left( t\right) =1-\mathop{\rm tr}\hat{\rho }%
_{a}^{2}\left( t\right) =\frac{1}{2}\left\{ 1-\exp \left\{ 2\mathop{\rm Re}%
\nolimits\Phi \left( \omega ,k,F;t\right) -D^{2}\left[ \beta _{e}^{\prime
}\left( t\right) ,\beta _{g}^{\prime }\left( t\right) \right] \right\}
\right\} .  \label{lin_entA}
\end{equation}
We can recognize two distinct contributions to the coherence loss of the
atomic state. These contributions are identified by the two terms in the
argument of the exponential in Eq. (\ref{lin_entA}). The first term,
proportional to the real part of the complex phase $\Phi $, reflects the
presence of dissipation and the external source. The second term is
proportional to the distance in the phase space between the states $\left|
\beta _{e}^{\prime }\left( t\right) \right\rangle $ and $\left| \beta
_{g}^{\prime }\left( t\right) \right\rangle $. Hence, the coherence
properties of the atomic state are affected by the presence of the thermal
reservoir and the external source, even if the atom is not directly coupled
to them, and by the entanglement process between atom and field.

Tracing out the global density operator in the atomic variables, we get the
reduced field density operator 
\begin{equation}
\hat{\rho }_{f}\left( t\right) =\frac{1}{2}\left\{ \left| \beta
_{e}^{\prime }\left( t\right) \right\rangle \left\langle \beta _{e}^{\prime
}\left( t\right) \right| +\left| \beta _{g}^{\prime }\left( t\right)
\right\rangle \left\langle \beta _{g}^{\prime }\left( t\right) \right|
\right\} ,  \label{rhoF3}
\end{equation}
whose idempotency defect is 
\begin{equation}
\varsigma _{f}\left( t\right) =1-\mathop{\rm tr}\hat{\rho }%
_{f}^{2}\left( t\right) =\frac{1}{2}\left\{ 1-\exp \left\{ -D^{2}\left[
\beta _{e}^{\prime }\left( t\right) ,\beta _{g}^{\prime }\left( t\right) %
\right] \right\} \right\} .  \label{lin_entF}
\end{equation}
The field state is a statistical mixture of the coherent states $\left|
\beta _{e}^{\prime }\left( t\right) \right\rangle $ and $\left| \beta
_{g}^{\prime }\left( t\right) \right\rangle $. Although the amplitudes $%
\beta _{e}^{\prime }\left( t\right) $ and $\beta _{g}^{\prime }\left(
t\right) $ have equal moduli, the phase between them varies in time in a
complicated form. Here, the contribution to the idempotency defect is due to
the entanglement process between atom and field. The asymptotic field state
is not a pure state but a statistical mixture of the coherent states $\left|
\beta _{e}^{\prime }\left( \infty \right) \right\rangle $ and $\left| \beta
_{g}^{\prime }\left( \infty \right) \right\rangle $, i.e., 
\begin{equation*}
\hat{\rho }_{f}\left( \infty \right) =\frac{1}{2}\left\{ \left| \beta
_{e}^{\prime }\left( \infty \right) \right\rangle \left\langle \beta
_{e}^{\prime }\left( \infty \right) \right| +\left| \beta _{g}^{\prime
}\left( \infty \right) \right\rangle \left\langle \beta _{g}^{\prime }\left(
\infty \right) \right| \right\} .
\end{equation*}

The operator $\hat{\rho }_{f}\left( t\right) $ can be diagonalized,
yielding the following eigenvectors and eigenvalues 
\begin{subequations}
\label{EV_field}
\begin{align}
\left| \varphi _{+}\right\rangle &= \frac{1}{N_{+}\left( t\right) }\left[
\left| \beta _{e}^{\prime }\left( t\right) \right\rangle +e^{i\chi \left(
t\right) }\left| \beta _{g}^{\prime }\left( t\right) \right\rangle \right] ,
\label{EV1_field} \\
\left| \varphi _{-}\right\rangle &= \frac{1}{N_{-}\left( t\right) }\left[
\left| \beta _{e}^{\prime }\left( t\right) \right\rangle -e^{i\chi \left(
t\right) }\left| \beta _{g}^{\prime }\left( t\right) \right\rangle \right] ,
\label{EV2_field}
\end{align}
\end{subequations}
\begin{subequations}
\label{ev_field}
\begin{align}
\Lambda _{+}\left( t\right) &= \frac{N_{+}^{2}\left( t\right) }{4}=\frac{1}{2%
}\left\{ 1+\exp \left\{ -\frac{1}{2}D^{2}\left[ \beta _{e}^{\prime }\left(
t\right) ,\beta _{g}^{\prime }\left( t\right) \right] \right\} \right\} ,
\label{ev1_field} \\
\Lambda _{-}\left( t\right) &= \frac{N_{-}^{2}\left( t\right) }{4}=\frac{1}{2%
}\left\{ 1-\exp \left\{ -\frac{1}{2}D^{2}\left[ \beta _{e}^{\prime }\left(
t\right) ,\beta _{g}^{\prime }\left( t\right) \right] \right\} \right\} ,
\label{ev2_field}
\end{align}
\end{subequations}
where $\chi \left( t\right) =\mathop{\rm Im}\nolimits\left[ \beta
_{e}^{\prime }\left( t\right) \beta _{g}^{\prime \ast }\left( t\right) %
\right] $.

\subsubsection{Short and long times behavior of the coherence loss of 
subsystems states}

For long times ($kt\gg 1$), the coherence loss of the atomic subsystem is
dominated by the linear dependence on $t$ of the function $\Phi $. In this
regime, the time dependence of the atomic linear entropy $\varsigma _{a}$
closely follows that for the global system $\varsigma $. Hence, the
contribution to the atomic coherence loss, for long times, is due to the
presence of the dissipation and the external source, and the time scales of
the atomic and the full system's decoherence are the same. On the other
hand, for short times ($kt\ll 1$ and $\omega t\ll 1$), the contribution for
the atomic coherence loss is solely due to the entanglement process. In
fact, in this regime, the dominant term in the argument of the exponential
in the RHS of Eq. (\ref{lin_entA}) is quadratic in $t$ and comes from the
factor proportional to the distance between the states $\left| \beta
_{e}^{\prime }\left( t\right) \right\rangle $ and $\left| \beta _{g}^{\prime
}\left( t\right) \right\rangle $, namely $D^{2}\left[ \beta _{e}^{\prime
}\left( t\right) ,\beta _{g}^{\prime }\left( t\right) \right] $. The atomic
linear entropy grows accordingly $\varsigma _{a}\sim 1-\exp \left[ -\left(
t/\tau _{a,\mathrm{dec}}^{\mathrm{st}}\right) ^{2}\right] $, where 
\begin{equation}
\tau _{a,\mathrm{dec}}^{\mathrm{st}}=\frac{k}{2\left| F\right| \omega }.
\label{st_charA}
\end{equation}
Since the decoherence of the atomic state for short times is due to the
entanglement process, it is faster in the subcritical damping ($k/\omega <1$%
). Moreover, increasing of the intensity of the source shortens the
atomic decoherence characteristic time.

The decoherence of the field state, for short times, is similar to that of
the atomic state. In fact, at $t\rightarrow 0$, the purity loss of both
systems is mainly due to the initial entanglement process. Hence, for short
times, the characteristic decoherence times of atom ($\tau _{a,\mathrm{dec}%
}^{\mathrm{st}}$) and field ($\tau _{f,\mathrm{dec}}^{\mathrm{st}}$) are
equal and related to the unitary interaction.

\subsection{Results and discussion}

As we pointed out, the linear term in $t$ appearing in the function $\Phi $
is responsible by the complete vanishing of the ``non-diagonal elements'' $%
\hat{\rho }_{eg}$ and $\hat{\rho }_{ge}$. Contrary to the model
studied in Ref. \cite{faria} -- the dispersive JCM with dissipation but
without coupling to an external source -- the decoherence of the full system
atom-field prepared in the initial state (\ref{initial_state}) is complete.
The graphs presented in Figs. 1 and 2 clearly exhibit this
behavior. In Fig. 1, the graphs of $\varsigma $, $\varsigma _{a}$
and $\varsigma _{f}$ as a function of $\omega t/\pi $ for different values of $%
k/\omega $ are shown, with constant $\left| F\right| /k$  ratio.
In Fig. 2 are plotted the graphs for two different values of the
ratio $\left| F\right| /k$ in the subcritical regime.

It is interesting to note that the larger the coupling between field and
external source, $\left| F\right| $, the more rapid the coherence loss of
both the full system atom-field and the atom only. Since the intensity of
the injected field \ by the source is a measure of its ``classicality'', the
coherence loss of the global system becomes faster as the intensity
increases. Moreover, the characteristic decoherence time $\tau _{\mathrm{dec}%
}^{\mathrm{lt}}$ is inversely proportional to mean number of photons in the
asymptotic state, $\overline{n}\left( \infty \right) =\left| F\right|
^{2}/\left( k^{2}+\omega ^{2}\right) $.

The dependence of $\varsigma \left( t\right) $ with the ratio $k/\omega $ is
more complicated. Since the characteristic decoherence time directly depends
on the characteristic dissipation time, $1/k$, one would expect
the decoherence
of the global system to be slower in a less dissipative environment. This
conjecture seems to be verified if one compares, in Fig. 1, the
curves of $\varsigma \left( t\right) $ for $k/\omega =0.2$ and $k/\omega =1$%
. Note that $\varsigma $ reaches the saturation rapidly in the case $%
k/\omega =1$. Hence, we expect that the larger the value of $k/\omega $, the
faster the saturation of $\varsigma $. But it does not happen: as shown in
Fig. 1, for $k/\omega =1$, $\varsigma $ reaches the plateau at $%
t=\pi /\omega $, while for $k/\omega =5$, the plateau is reached at $t=3\pi
/2\omega $, approximately. Roughly, the coherent superposition between the
states $\left| e,\beta _{e}^{\prime }\left( t\right) \right\rangle $ and $%
\left| g,\beta _{g}^{\prime }\left( t\right) \right\rangle $ that the
unitary contribution tries to create, is transformed by non-unitary mechanism
into a statistical mixture. The more distinguishable the states that form the
superposition, the faster the coherence loss. A measure of
distinguishability is provided by the distance in the phase space between
the states $\left| \beta _{e}^{\prime }\left( t\right) \right\rangle $ and $%
\left| \beta _{g}^{\prime }\left( t\right) \right\rangle $. The amplitudes $%
\beta _{e}^{\prime }\left( t\right) $ and $\beta _{g}^{\prime }\left(
t\right) $ are inversely proportional to $\omega \pm ik$; hence, on the one
hand, if the increasing of $k$ favours decoherence, on the other hand, high
dissipation taxes rapidly diminishes the separability between the states $%
\left| \beta _{e}^{\prime }\left( t\right) \right\rangle $ and $\left| \beta
_{g}^{\prime }\left( t\right) \right\rangle $ and the decoherence becomes
slower.

As the model studied in Ref. \cite{faria}, the atom is more influenced by
non-unitary dynamics. In fact, the contribution to the purity loss of the
atomic state results both from the interaction between atom and field and
from the presence of the dissipative environment. These different
contributions can be identified by the terms proportional to $\mathop{\rm Re}%
\nolimits\Phi \left( \omega ,k,F;t\right) $ and to $D^{2}\left[ \beta
_{e}^{\prime }\left( t\right) ,\beta _{g}^{\prime }\left( t\right) \right] $
in the argument of the exponential in Eq. (\ref{lin_entA}). Hence, the
purity loss of the atomic state is complete, as one can verify in the graphs
in Fig. 1, specially for the cases $k/\omega =1$ and $k/\omega =5$.
On the other hand, the decoherence of the field state only results from the
interaction between atom and field. However, contrary to the model studied
in Ref. \cite{faria}, the asymptotic state of the field is formed by a
statistical mixture of the coherent states $\left| \beta _{e}^{\prime
}\left( \infty \right) \right\rangle $ and $\left| \beta _{g}^{\prime
}\left( \infty \right) \right\rangle $. The asymptotic value of the linear
entropy $\varsigma _{f}$ is larger in the critical regime ($k/\omega =1$)
than in the subcritical and supercritical regimes.

It is worth to note that the linear entropy of the field $\varsigma _{f}$
exhibits local maxima and minima, specially in the subcritical regime. These
local maxima and minima correspond to the instants ($t_{c}$) of maximum and
minimum distance between the states $\left| \beta _{e}^{\prime }\left(
t\right) \right\rangle $ and $\left| \beta _{g}^{\prime }\left( t\right)
\right\rangle $. These critical instants can be calculated by the zeros of
the time derivative of 
\begin{equation}
D^{2}\left[ \beta _{e}^{\prime }\left( t\right) ,\beta _{g}^{\prime }\left(
t\right) \right] =\frac{4\left| F\right| ^{2}\omega ^{2}}{k^{2}\left(
k^{2}+\omega ^{2}\right) }\left[ k\left( e^{-kt}\cos \omega t-1\right)
-\omega e^{-kt}\mathop{\rm sin}\nolimits\omega t\right] ^{2}.
\label{distance}
\end{equation}
In this way, the critical instants shall satisfy 
\begin{equation}
k\left( e^{-kt_{c}}\cos \omega t_{c}-1\right) -\omega e^{-kt_{c}}\mathop{\rm
sin}\nolimits\omega t_{c}=0  \label{tc1}
\end{equation}
or 
\begin{equation}
\cos \omega t_{c}=0.  \label{tc2}
\end{equation}
When $t_{c}$ satisfies (\ref{tc1}), the distance in (\ref{distance})
vanishes, i.e., $\beta _{e}^{\prime }\left( t_{c}\right) =\beta _{g}^{\prime
}\left( t_{c}\right) $. In this case, $\varsigma _{f}$ is null, the field is
found in a pure state and the global state disentangles. On the other hand,
if $\omega t_{c}=\left( 2n+1\right) \pi /2$, $n$ integer, Eq. (\ref{tc2}) is
satisfied. Now, we can have a local maximum or a local minimum depending of
the signal of the second time derivative of $D^{2}\left[ \beta _{e}^{\prime
}\left( t\right) ,\beta _{g}^{\prime }\left( t\right) \right] $. At the
critical instants, we have 
\begin{equation*}
\mathop{\rm sgn}\left\{ \left. \frac{d^{2}}{dt^{2}}D^{2}
\left[ \beta _{e}^{\prime}\left( t\right) ,
\beta _{g}^{\prime }\left( t\right) \right]\right|_{t=t_{c}}
\right\} =\mathop{\rm sgn}
\left[ \sin \omega t_{c}\left( k+\omega
e^{-kt_{c}}\sin \omega t_{c}\right) \right] .
\end{equation*}
We consider two different cases:

\begin{enumerate}
\item  Critical and supercritical regimes ($k/\omega \geq 1$): If $\omega
t_{c}=\left( 2n+1\right) \pi /2$, $n$ even, we have a local maximum,
otherwise, we have a local minimum, but the value of $\varsigma _{f}$ is not
null, i.e., the field state is characterized by a statistical mixture.

\item  Subcritical regime ($k/\omega <1$): If $\omega t_{c}=\left(
2n+1\right) \pi /2$, $n$ even, we have a local maximum. On the other hand,
if $n$ is odd, in order the signal of the second time derivative to be
positive, $t_{c}$ shall satisfy 
\begin{equation*}
t_{c}>\frac{1}{k}\ln \frac{\omega }{k}=t_{\mathrm{trans}}.
\end{equation*}
Hence, if $t_{c}=\left( 2n+1\right) \pi /2\omega <t_{\mathrm{trans}}$, $n$
odd, we have a local maximum, otherwise, if $t_{c}>t_{\mathrm{trans}}$, we
have a local minimum. Note that the local minima corresponding to $t_{c}>t_{%
\mathrm{trans}}$, $\varsigma _{f}$ is not null.
\end{enumerate}

\subsection{Measuring correlations}

According to Werner \cite{separability}, the density operator $\hat{\rho }$
which represents the state of a bipartite system $A+B$ is said
disentangled (or separable) iff 
\begin{equation}
\hat{\rho }=\sum_{i}p_{i}\hat{\rho }_{A}^{i}\otimes \hat{\rho }%
_{B}^{i},  \label{classical}
\end{equation}
where $\hat{\rho }_{A}^{i}$($\hat{\rho }_{B}^{i}$) are density
operators on the state space of the system $A$($B$). $p_{i}$ are
non-negative real numbers, such as $\sum_{i}p_{i}=1$. If $\hat{\rho }$
cannot be written in form (\ref{classical}), the state is said entangled or
quantum correlated. Moreover, we can demand both $\hat{\rho }_{A}^{i}$
and $\hat{\rho }_{B}^{i}$ to be a pure state.

The state of each subsystem is described by the reduced density operators, $%
\hat{\rho }_{A}\equiv \mathop{\rm tr}_{B}\hat{\rho }$ and $\hat{%
\rho }_{B}\equiv \mathop{\rm tr}_{A}\hat{\rho }$. However, in general,
the global state cannot be determined from the reduced states. In short, $%
\hat{\rho }\neq \hat{\rho }_{A}\otimes \hat{\rho }_{B}$. An
important information about the global state is lost in the partial
tracing out procedure. This information is related to the local
(classical) and
non-local (quantum) correlations between the two subsystems, $A$ and $B$. We
can ask about the ``distance'' between the global state $\hat{\rho }$
and the corresponding completely uncorrelated state $\hat{\rho }%
_{A}\otimes \hat{\rho }_{B}$ as a measure of the total correlation of
the state $\hat{\rho }$. A possible choice of distance is given by the
Hilbert-Schmidt metric, hence, the total correlation measure of the state $%
\hat{\rho }$ is defined as 
\begin{equation*}
c\left( \hat{\rho }\right) \equiv d^{2}\left( \hat{\rho },\hat{%
\rho }_{A}\otimes \hat{\rho }_{B}\right) =\left| \left| \hat{\rho }-%
\hat{\rho }_{A}\otimes \hat{\rho }_{B}\right| \right| ^{2}\equiv %
\mathop{\rm tr}\left( \hat{\rho }-\hat{\rho }_{A}\otimes \hat{%
\rho }_{B}\right) ^{2}.
\end{equation*}
Returning to dispersive JCM, the expression for $c\left( \hat{\rho }%
\right) $ can be written as 
\begin{equation}
c\left( \hat{\rho }\right) =\frac{\varsigma _{f}\left( t\right) }{2}%
\left\{ 1+\left[ 1-2\varsigma \left( t\right) \right] \left[ 1+2\varsigma
_{f}\left( t\right) \right] \right\} .  \label{total_corr}
\end{equation}
$c\left( \hat{\rho }\right) $ is a non-negative quantity; if the field
is found in a pure state, we have $c\left( \hat{\rho }\right) =0$ and
the global state is characterized by a completely uncorrelated state.

The correlation measure above defined does not distinguish classical and
quantum correlations. In order to evaluate the entanglement of the system
atom-field, we chose the concurrence \cite{concurrence} as measure of degree
of entanglement. It has been proven to be
a reasonable entanglement measure for mixed
states of bipartite systems composed by two-level subsystems. Since the
reduced state of the field has rank no greater than two, we can effectively
consider the global system atom-field formed by two two-level subsystems at
each instant of time $t$. If the density matrix $\hat{\rho }$ represents
the state of two two-level systems $A$ and $B$, the concurrence is defined
as 
\begin{equation*}
C\left( \hat{\rho }\right) =\max \left\{
0,x_{1}-x_{2}-x_{3}-x_{4}\right\} ,
\end{equation*}
where $x_{1}\geq x_{2}\geq x_{3}\geq x_{4}$ are the eigenvalues of the
matrix $\sqrt{\hat{\rho }^{1/2}\widetilde{\rho }\hat{\rho }^{1/2}}$.
The matrix $\widetilde{\rho }$ is given by 
\begin{equation*}
\widetilde{\rho }=\left( \hat{\sigma }_{y}\otimes \hat{\sigma }%
_{y}\right) \hat{\rho }^{\ast }\left( \hat{\sigma }_{y}\otimes 
\hat{\sigma }_{y}\right) .
\end{equation*}
Here, $\hat{\rho }^{\ast }$ represents the complex conjugation of $%
\hat{\rho }$ in a fixed basis. $\hat{\sigma }_{y}$ is the Pauli
pseudo-matrix 
\begin{equation*}
\hat{\sigma }_{y}=\left( 
\begin{array}{cc}
0 & -i \\ 
i & 0
\end{array}
\right)
\end{equation*}
in the same basis. Note that $0\leq C\left( \hat{\rho }\right) \leq 1$.
The upper limit indicates maximum entanglement; the lower limit is
characteristic of separable states.

The concurrence of the global atom-field state is given by the expression 
\begin{align}
C\left( \hat{\rho }\right) &= 2\left| \lambda _{+}\left( t\right)
-\lambda _{-}\left( t\right) \right| \sqrt{\Lambda _{+}\left( t\right)
\Lambda _{-}\left( t\right) }  \label{concur} \\
&= 2\exp \left[ \Phi \left( \omega ,k,F;t\right) \right] \left\{ 1-\exp
\left\{ -D^{2}\left[ \beta _{e}^{\prime }\left( t\right) ,\beta _{g}^{\prime
}\left( t\right) \right] \right\} \right\} ^{1/2},  \notag
\end{align}
where $\lambda _{\pm }\left( t\right) $ and $\Lambda _{\pm }\left( t\right) $
are the eigenvalues of the global system and of the field, respectively. The
graphs of the correlation measure $c\left( \hat{\rho }\right) $\ and the
concurrence $C\left( \hat{\rho }\right) $\ as a function of time for the
three dynamical regimes are displayed in Fig. 3. As expected, the
asymptotic value of the correlation measure $c\left( \hat{\rho }\right) $
is not null, since the global system evolves to a classically correlated
state. Hence, 
\begin{equation*}
c\left[ \hat{\rho }\left( \infty \right) \right] =\frac{\varsigma
_{f}\left( \infty \right) }{2}.
\end{equation*}
Keeping the value of $\left| F\right| /k$ constant, the maximum value of $c%
\left[ \hat{\rho }\left( \infty \right) \right] $ occurs in the critical
regime.

In the three dynamical regimes, the concurrence vanishes in the asymptotic
limit (this is noticeable in critical and supercritical regimes). In fact,
the non-unitary mechanism completely destroys any trace of entanglement
between atom and field, despite of the continuous pumping of the field by
the external source. Since the global state evolves to a complete
statistical mixture, both the eigenvalues of the density operator $\hat{%
\rho }$, $\lambda _{+}$ and $\lambda _{-}$, tend to be equal to $1/2$.
Hence, the long times behavior of $C\left( \hat{\rho }\right) $ is
mainly governed by the factor $\left| \lambda _{+}\left( t\right) -\lambda
_{-}\left( t\right) \right| $ in Eq. (\ref{concur}). Note that the decay of
the concurrence in the critical regime is more rapid than the corresponding
decay in the supercritical regime. In the subcritical regime, at the
instants when atom and field are disentangled, $C\left( \hat{\rho }%
\right) $ is null, as expected. In these instants, the field is found in a
pure state, $\hat{\rho }_{f}$ has rank equal to unit and $\det \hat{%
\rho }_{f}\left( t\right) =\Lambda _{+}\left( t\right) \Lambda _{-}\left(
t\right) $ disappears. We can conclude that the degree of entanglement
between atom and field results from the competition of two processes: the
unitary interaction between them and the dissipative dynamics due to the
coupling between field and environment. The contribution of the unitary
process for the concurrence can be recognized by the presence of the square
root of $\Lambda _{+}\left( t\right) \Lambda _{-}\left( t\right) $ in Eq. (%
\ref{concur}), whereas the effects of the non-unitary mechanism are carried
in factor $\left| \lambda _{+}\left( t\right) -\lambda _{-}\left( t\right)
\right| $. These statements remain true even the source is eliminated. In
this case, the field evolves to the vacuum state and the asymptotic global
state is completely uncorrelated \cite{faria}.

\appendix

\section{On the solutions of the equations of motion of the ``matrix
elements'' of global density operator}

As we discussed above, the solutions of the equations of motion (\ref
{gen_form}) of the ``matrix elements'' $\hat{\rho }_{ij}=\left\langle
i\right| \hat{\rho }\left( t\right) \left| j\right\rangle $, $i,j=\left(
e,g\right) $, are given by the general formula (\ref{formal2}). Each
dynamical generator $\mathcal{L}_{ij}^{\prime }$ is a linear combination of
elements of some Lie algebra and the action of the exponential $\exp \left( 
\mathcal{L}_{ij}^{\prime }t\right) $ (so-called ``Lie exponential") on the
initial state $\hat{\rho }_{ij}\left( 0\right) $ might be easily
evaluated if one expresses it as an ordered product of exponentials of
elements of the corresponding algebra \cite{wilcox,lie_algebra,witschel}.
We obtain
the suitable similarity transformation of the Lie exponentials involved in
the solutions of \ the equations of motion for $\hat{\rho }_{ee}$, etc.,
by using the technique developed by Wilcox \cite{wilcox} known as parameter
differentiation method.

\subsection{The parameter differentiation method}

The parameter differentiation method \cite{wilcox} uses the Baker-Hausdorff
formula to expand a Lie exponential in an ordered product of exponentials.
If $\hat{A}$ and $\hat{B}$ are two operators that do not commute,
the expression 
\begin{equation}
e^{\hat{A}}\hat{B}e^{-\hat{A}}=\hat{B}+\left[ \hat{A},%
\hat{B}\right] +\frac{1}{2!}\left[ \hat{A},\left[ \hat{A},%
\hat{B}\right] \right] +\cdots +\frac{1}{n!}\underbrace{\left[ \hat{A},%
\left[ \hat{A},\cdots \left[ \hat{A},\hat{B}\right] \cdots \right] %
\right] }_{n\ \mathrm{nested\ commutators}}+\cdots   \label{BH_formula1}
\end{equation}
is known as Baker-Hausdorff formula. In short, it can be rewritten as 
\begin{equation}
e^{\hat{A}}\hat{B}e^{-\hat{A}}=e^{\left[ \hat{A},\cdot %
\right] }\hat{B}=\sum_{n=0}^{\infty }\frac{1}{n!}\left[ \hat{A}%
,\cdot \right] ^{n}\hat{B}.  \label{BH_formula2}
\end{equation}
The superoperator $\left[ \hat{A},\cdot \right] ^{n}$ represents the
recurrent application of the commutator $\left[ \hat{A},\cdot \right] $, 
\begin{gather*}
\left[ \hat{A},\cdot \right] ^{n} =\left[ \hat{A},\cdot \right]
^{n-1}\left[ \hat{A},\cdot \right] , \\
\left[ \hat{A},\cdot \right] ^{1} =\left[ \hat{A},\cdot \right] ,
\\
\left[ \hat{A},\cdot \right] ^{0} = 1
\end{gather*}
and

\begin{equation*}
\left[ \hat{A},\cdot \right] \hat{B}=\left[ \hat{A},\hat{B}%
\right] .
\end{equation*}

Let us consider the $n$-dimensional Lie algebra $\mathcal{A}_{n}=\left\{ 
\hat{A}_{1},\hat{A}_{2},\cdots ,\hat{A}_{n}\right\} $, where the
commutator between any pair $\hat{A}_{i},\hat{A}_{j}\in \mathcal{A}%
_{n}$ is expressed as a linear combination of elements of $\mathcal{A}_{n}$%
, i.e., 
\begin{equation}
\left[ \hat{A}_{i},\hat{A}_{j}\right] =\sum_{k=1}^{n}C_{i,j}^{k}%
\hat{A}_{k}.  \label{comm1}
\end{equation}
The coefficients $\left\{ C_{i,j}^{k}\right\} $ are real or complex numbers
called structure constants of the algebra $\mathcal{A}_{n}$ \cite
{lie_algebra}. We define a Lie exponential as an exponential of any linear
combination $\mathcal{L}=a_{1}\hat{A}_{1}+a_{2}\hat{A}_{2}+\cdots +a_{n}%
\hat{A}_{n}=\sum_{i=1}^{n}a_{i}\hat{A}_{i}$ of elements of $\mathcal{%
A}_{n}$, i.e., 
\begin{equation}
\exp \left( \mathcal{L}t\right) \equiv \exp \left\{ \left( a_{1}\hat{A}%
_{1}+a_{2}\hat{A}_{2}+\cdots +a_{n}\hat{A}_{n}\right) t\right\} ,
\label{Lie_exp1}
\end{equation}
where $\left\{ a_{i}\right\} _{i=1,\ldots ,n}$ are the coefficients of the
linear combination and $t$ is a real or complex parameter. The exponential (%
\ref{Lie_exp1}) can be expressed as an ordered product of exponentials, 
\begin{equation}
\exp \left( \mathcal{L}t\right) =\exp \left\{ f_{1}\left(
a_{1},\cdots ,a_{n};t\right) \hat{A}_{1}\right\} \times \cdots \times \exp
\left\{ f_{n}\left( a_{1},\cdots ,a_{n};t\right) \hat{A}_{n}\right\} .
\label{Lie_exp2}
\end{equation}
Here, $f_{i}\left( a_{1},\cdots ,a_{n};t\right) $, $i=1,\ldots ,n$, is a
function of the coefficients $\left\{ a_{i}\right\} _{i=1,\ldots ,n}$ and of
the parameter $t$. The parameter differentiation method allow us to
determine these functions.

Differentiating both sides of (\ref{Lie_exp2}) with respect to the parameter 
$t$, we get 
\begin{align}
\frac{d}{dt}\exp \left( \mathcal{L}t\right) &=\mathcal{L}\exp \left( 
\mathcal{L}t\right)  \notag \\
&= \overset{\cdot }{f}_{1}\left( a_{1},\cdots ,a_{n};t\right) \hat{A}%
_{1}\exp \left\{ f_{1}\left( a_{1},\cdots,a_{n};t\right) \hat{A}_{1}\right\}
\notag \\
& \times \cdots \times \exp \left\{ f_{n}\left( a_{1},\cdots ,a_{n};t\right) 
\hat{A}_{n}\right\}  \notag \\
& +\cdots +\overset{\cdot }{f}_{n}\left( a_{1},\cdots ,a_{n};t\right) 
\exp \left\{
f_{1}\left( a_{1},\cdots ,a_{n};t\right) \hat{A}_{1}\right\}
\label{derivative} \\
& \times \cdots \times \hat{A}_{n}\exp \left\{ f_{n}\left(
a_{1},\cdots ,a_{n};t\right) \hat{A}_{n}\right\}  \notag \\
& = \sum_{i=1}^{n}\overset{\cdot }{f}_{i}\left( a_{1},\cdots ,a_{n};t\right) 
\hat{B}_{i1}\exp \left\{ f_{1}\left( a_{1},\cdots ,a_{n};t\right) \hat{A}%
_{1}\right\}  \notag \\
& \times \cdots \times \hat{B}_{in}\exp \left\{ f_{n}\left(
a_{1},\cdots ,a_{n};t\right) \hat{A}_{n}\right\} .  \notag
\end{align}
Here, dot indicates derivatives with respect to $t$, and the operators $%
\hat{B}_{ij}$, $i,j=1,\cdots ,n$, are given by 
\begin{equation*}
\hat{B}_{ij}=\left\{ 
\begin{array}{c}
\hat{A}_{i},\ \mathrm{if\ }i=j \\ 
1,\ \mathrm{if\ }i\neq j
\end{array}
\right. .
\end{equation*}
The repeated application of the Baker-Hausdorff formula (\ref{BH_formula1})
allows us to move the operators $\hat{A}_{i}$, $i=2,\cdots ,n$, to the left
position in each term of the sum appearing in (\ref{derivative}). For
instance, the second term of this sum is 
\begin{align*}
& \overset{\cdot }{f}_{2}\left( a_{1},\cdots ,a_{n};t\right) \exp \left\{
f_{1}\left( a_{1},\cdots ,a_{n};t\right) \hat{A}_{1}\right\} \hat{A}%
_{2}\exp \left\{ f_{2}\left( a_{1},\cdots ,a_{n};t\right) \hat{A}_{2}\right\}
\\
& \qquad \qquad \times \cdots \times \exp \left\{ f_{n}\left(
a_{1},\cdots ,a_{n};t\right) \hat{A}_{n}\right\} .
\end{align*}
To move the operator $\hat{A}_{2}$ to the left of the exponential in $%
\hat{A}_{1}$, we write 
\begin{align}
\exp \left\{ f_{1}\left( a_{1},\cdots ,a_{n};t\right) \hat{A}_{1}\right\} 
\hat{A}_{2} &= \exp \left\{ f_{1}\left( a_{1},\cdots ,a_{n};t\right) 
\hat{A}_{1}\right\} \hat{A}_{2}  \notag \\
& \times \exp \left\{ -f_{1}\left( a_{1},\cdots ,a_{n};t\right) \hat{A}%
_{1}\right\}  \label{dummy} \\
& \times \exp \left\{ f_{1}\left( a_{1},\cdots ,a_{n};t\right) \hat{A}%
_{1}\right\} .  \notag
\end{align}
The Baker-Hausdorff formula yields 
\begin{equation*}
\exp \left\{ f_{1}\left( a_{1},\cdots ,a_{n};t\right) \hat{A}_{1}\right\} 
\hat{A}_{2}\exp \left\{ -f_{1}\left( a_{1},\cdots ,a_{n};t\right) \hat{A}%
_{1}\right\} =\mathbb{G}_{2}\left( \hat{A}_{1},\cdots ,\hat{A}%
_{n};f_{1}\right) ,
\end{equation*}
where $\mathbb{G}_{2}$ is a function of $\hat{A}_{1},\cdots ,\hat{A}_{n}$
and $f_{1}$. If this procedure is performed on the other terms of the sum in
(\ref{derivative}), we find 
\begin{equation}
\mathcal{L}=\sum_{i=1}^{n}\overset{\cdot }{f}_{i}\left(
a_{1},\cdots ,a_{n};t\right) \mathbb{G}_{i}\left( \hat{A}_{1},\cdots ,\hat{A}%
_{n};f_{1,\cdots ,}f_{i-1}\right) .  \label{identity1}
\end{equation}
The functions $\mathbb{G}_{i}$ are given by 
\begin{align*}
\mathbb{G}_{1}\left( \hat{A}_{1},\cdots ,\hat{A}_{n}\right) &= \hat{A%
}_{1}, \\
\mathbb{G}_{2}\left( \hat{A}_{1},\cdots ,\hat{A}_{n};f_{1}\right)
&= \exp \left\{ f_{1}\left( a_{1},\cdots ,a_{n};t\right) \hat{A}_{1}\right\} 
\hat{A}_{2} \\
& \times \exp \left\{ -f_{1}\left( a_{1},\cdots ,a_{n};t\right) \hat{A}%
_{1}\right\} , \\
\mathbb{G}_{3}\left( \hat{A}_{1},\cdots ,\hat{A}_{n};f_{1},f_{2}\right)
&= \exp \left\{ f_{1}\left( a_{1},\cdots ,a_{n};t\right) \hat{A}_{1}\right\}
\exp \left\{ f_{2}\left( a_{1},\cdots ,a_{n};t\right) \hat{A}_{2}\right\} \\
& \times \hat{A}_{3}\exp \left\{ -f_{2}\left( a_{1},\cdots ,a_{n};t\right) 
\hat{A}_{2}\right\} \\
& \times \exp \left\{ -f_{1}\left( a_{1},\cdots ,a_{n};t\right) \hat{A}%
_{1}\right\} ,
\end{align*}
and so on. The identity (\ref{identity1}) yields a set of coupled
differential equations for the functions $\left\{ f_{i}\right\} _{i=1,\ldots
,n}$. The solution of this system of differential equations with the
corresponding initial condition determines the functions $\left\{
f_{i}\right\} _{i=1,\ldots ,n}$.

\subsection{The algebra of the bosonic superoperators}

The dynamical generator $\mathcal{L}_{ij}^{\prime }$ which appears in the
general form of equation of motion (\ref{master3}) for the ``matrix
elements'' $\hat{\rho }_{ij}=\left\langle i\right| \hat{\rho }\left(
t\right) \left| j\right\rangle $, $i,j=\left( e,g\right) $, is a linear
combination of bosonic superopertors \cite{sup-op,faria}, which form a
finite Lie algebra under commutation. The bosonic superoperators represent
the action of creation and annihilation operators of the harmonic
oscillator, $\hat{a}^{\dagger }$ and $\hat{a}$, on an operator $%
\hat{O}$: 
\begin{equation}
a^{l}\hat{O}=\hat{a}\hat{O},\ a^{l\dagger }\hat{O}=\hat{a%
}^{\dagger }\hat{O},\ a^{r}\hat{O}=\hat{O}\hat{a},\
a^{r\dagger }\hat{O}=\hat{O}\hat{a}^{\dagger }.
\label{boson_supop}
\end{equation}
The sets $\left\{ a^{l},a^{l\dagger },1\right\} $ and $\left\{
a^{r},a^{r\dagger },1\right\} $ constitute left and right realization of
the Heisenberg-Weyl group $hw\left( 4\right) $ \cite{perelomov}, denoted $%
hw_{l}\left( 4\right) $ and $hw_{r}\left( 4\right) $, respectively. From the
fundamental relation $\left[ \hat{a},\hat{a}^{\dagger }\right] =1$
and the above definitions, we derive the commutation relations between the
bosonic superoperators: 
\begin{gather}
\left[ a^{l},a^{l\dagger }\right]  = 1,  \label{comm2} \\
\left[ a^{r},a^{r\dagger }\right]  = -1.  \notag
\end{gather}
An superoperator belonging to $hw_{l}\left( 4\right) $ commutes with another
belonging to $hw_{r}\left( 4\right) $. The bilinear products of these
superoperators are 
\begin{gather}
\mathcal{M} \equiv a^{l\dagger }a^{l},  \notag \\
\mathcal{P} \equiv a^{r}a^{r\dagger },  \label{bilinear} \\
\mathcal{J} \equiv a^{l}a^{r\dagger }=a^{r\dagger }a^{l}.  \notag
\end{gather}
By virtue of the presence of the unitary term $\mathcal{L}_{S}$ 
[\textit{cf.} Eq. (\ref{liouvillianS})] due to the external source, 
it is convenient to define the following superoperators 
\begin{align}
\mathcal{X}_{\pm } &\equiv a^{l\dagger }\pm a^{r\dagger },  \label{sup_op1}
\\
\mathcal{Y}_{\pm } &\equiv a^{r}\pm a^{l}.  \notag
\end{align}
The superoperators above defined generate a finite Lie algebra. The non-null
commutation relations between these superoperators are given by 
\begin{gather}
\left[ \mathcal{J},\mathcal{M}\right] = \mathcal{J},  \notag \\
\left[ \mathcal{J},\mathcal{P}\right] = \mathcal{P},  \notag \\
\left[ \mathcal{J},\mathcal{X}_{\pm }\right] = \frac{1}{2}\left( \mathcal{X}%
_{+}-\mathcal{X}_{-}\right) ,  \notag \\
\left[ \mathcal{J},\mathcal{Y}_{\pm }\right] = \frac{1}{2}\left( \mathcal{Y}%
_{+}-\mathcal{Y}_{-}\right) ,  \notag \\
\left[ \mathcal{M},\mathcal{X}_{\pm }\right] = \frac{1}{2}\left( \mathcal{X}%
_{+}+\mathcal{X}_{-}\right) ,  \label{comm3} \\
\left[ \mathcal{M},\mathcal{Y}_{\pm }\right] = \pm \frac{1}{2}\left( 
\mathcal{Y}_{-}-\mathcal{Y}_{+}\right) ,  \notag \\
\left[ \mathcal{P},\mathcal{X}_{\pm }\right] = \pm \frac{1}{2}\left( 
\mathcal{X}_{-}-\mathcal{X}_{+}\right) ,  \notag \\
\left[ \mathcal{P},\mathcal{Y}_{\pm }\right] = \frac{1}{2}\left( \mathcal{Y}%
_{+}+\mathcal{Y}_{-}\right) ,  \notag \\
\left[ \mathcal{X}_{+},\mathcal{Y}_{-}\right] = -\left[ \mathcal{X}_{-},%
\mathcal{Y}_{+}\right] =2.  \notag
\end{gather}

\subsection{On the disentanglement of the Lie exponential corresponding to
the ``diagonal elements''}

The solution of the equation of motion for the operator $\hat{\rho }%
_{ee} $ is given by 
\begin{equation*}
\hat{\rho }_{ee}\left( t\right) =\exp \left( \mathcal{L}_{ee}^{\prime
}t\right) \hat{\rho }_{ee}\left( 0\right) .
\end{equation*}
In terms of the superoperators above defined, the Liouvillian $\mathcal{L}%
_{ee}^{\prime }$ can be expressed as 
\begin{equation*}
\mathcal{L}_{ee}^{\prime }=-i\omega \left( \mathcal{M}-\mathcal{P}\right)
+k\left( 2\mathcal{J}-\mathcal{M}-\mathcal{P}\right) -i\left( F\mathcal{X}%
_{-}-F^{\ast }\mathcal{Y}_{-}\right) =\mathcal{L}_{+}+\mathcal{L}_{S},
\end{equation*}
where $\mathcal{L}_{+}$ is given in Eq. (\ref{liouvillians}). From Eq. (\ref
{comm2}), we found 
\begin{align*}
\left[ \mathcal{L}_{+},\mathcal{X}_{-}\right] &=-\left( k+i\omega \right) 
\mathcal{X}_{-}, \\
\left[ \mathcal{L}_{+},\mathcal{Y}_{-}\right] &=-\left( k-i\omega \right) 
\mathcal{Y}_{-}.
\end{align*}
Hence, the algebra generated by the set $\left\{ \mathcal{L}_{+},\mathcal{X}%
_{-},\mathcal{Y}_{-}\right\} $ is an union of two two-dimensional Lie
subalgebras.

Let us express the Lie exponential $\exp \left( \mathcal{L}_{ee}^{\prime
}t\right) $ as 
\begin{equation}
\exp \left( \mathcal{L}_{ee}^{\prime }t\right) =\exp \left[ x_{-}\left(
t\right) \mathcal{X}_{-}\right] \exp \left[ y_{-}\left( t\right) \mathcal{Y}%
_{-}\right] \exp \left[ \lambda \left( t\right) \mathcal{L}_{+}\right] .
\label{Lie_exp3}
\end{equation}
We just have to determine the functions $x_{-}$, $y_{-}$, $\lambda $, given
the initial condition 
\begin{equation}
x_{-}\left( 0\right) =y_{-}\left( 0\right) =\lambda \left( 0\right) =0.
\label{initial_cond1}
\end{equation}
Taking the derivative of both sides of Eq. (\ref{Lie_exp3}) with respect to $%
t$, and applying the well-known result 
\begin{equation*}
e^{x\hat{A}}\hat{B}e^{-x\hat{A}}=e^{\beta x}\hat{B}+\left(
e^{\beta x}-1\right) \frac{\alpha }{\beta }
\end{equation*}
for a two-dimensional algebra $\mathcal{A}_{2}=\left\{ \hat{A},\hat{B%
}\right\} $, where $\left[ \hat{A},\hat{B}\right] =\alpha \hat{A}%
+\beta \hat{B}$, $\alpha ,\beta $ are c-numbers, we find 
\begin{equation*}
\mathcal{L}_{ee}^{\prime}=\left[ \overset{\cdot }{x}_{-}+\overset{\cdot }{%
\lambda }\left( k+i\omega \right) x_{-}\right] \mathcal{X}_{-}+\left[ 
\overset{\cdot }{y}_{-}+\overset{\cdot }{\lambda }\left( k-i\omega \right)
y_{-}\right] \mathcal{Y}_{-}+\overset{\cdot }{\lambda }\mathcal{L}_{+}.
\end{equation*}
This identity yields the following set of differential equations 
\begin{gather}
\overset{\cdot }{\lambda } = 1,  \notag \\
\overset{\cdot }{x}_{-}+\overset{\cdot }{\lambda }\left( k+i\omega \right)
x_{-} = -iF,  \label{diff_eq1} \\
\overset{\cdot }{y}_{-}+\overset{\cdot }{\lambda }\left( k-i\omega \right)
y_{-} = iF^{\ast }.  \notag
\end{gather}
Taking the initial conditions (\ref{initial_cond1}) into account, the
solution of these equations is 
\begin{gather}
\lambda \left( t\right) = t,  \notag \\
x_{-}\left( t\right) =\frac{F}{\omega -ik}\left[ e^{-\left( k+i\omega
\right) t}-1\right] = \beta _{e}\left( t\right) ,  \label{sol_diff_eq1} \\
y_{-}\left( t\right) = \frac{F^{\ast }}{\omega +ik}\left[ e^{-\left(
k-i\omega \right) t}-1\right] =\beta _{e}^{\ast }\left( t\right) .  \notag
\end{gather}
Note that 
\begin{equation*}
\exp \left( z\mathcal{X}_{-}\right) \exp \left( z^{\ast }\mathcal{Y}%
_{-}\right) =\exp \left( za^{l\dagger }-z^{\ast }a^{l}\right) \exp \left(
-za^{r\dagger }+z^{\ast }a^{r}\right) ,
\end{equation*}
therefore, we can express $\hat{\rho }_{ee}\left( t\right) $ as 
\begin{equation}
\hat{\rho }_{ee}\left( t\right) =\hat{D}\left[ \beta _{e}\left(
t\right) \right] \left[ \exp \left( \mathcal{L}_{+}t\right) \hat{\rho }%
_{ee}\left( 0\right) \right] \hat{D}^{\dagger }\left[ \beta _{e}\left(
t\right) \right] ,  \label{rhoee2}
\end{equation}
where $\hat{D}$ is the displacement operator of the Heisenberg-Weyl
group [\textit{cf.} Eq. (\ref{displacement})]. This procedure allows 
us to find an
analogous expression for the operator $\hat{\rho }_{gg}\left( t\right) $: 
\begin{equation*}
\hat{\rho }_{gg}\left( t\right) =\hat{D}\left[ \beta _{g}\left(
t\right) \right] \left[ \exp \left( \mathcal{L}_{-}t\right) \hat{\rho }%
_{gg}\left( 0\right) \right] \hat{D}^{\dagger }\left[ \beta _{g}\left(
t\right) \right] .
\end{equation*}
$\mathcal{L}_{-}$ and $\beta _{g}\left( t\right) $ are given by Eqs. (\ref
{liouvillians}) and (\ref{betag}), respectively.

\subsection{On the disentanglement of the Lie exponential corresponding to
the ``non-diagonal'' elements}

The solution of the equation of motion for the operator $\hat{\rho }%
_{eg} $ is given by 
\begin{equation*}
\hat{\rho }_{eg}\left( t\right) =\exp \left( \mathcal{L}_{eg}^{\prime
}t\right) \hat{\rho }_{eg}\left( 0\right) .
\end{equation*}
In terms of the bosonic superoperators, $\mathcal{L}_{eg}^{\prime }$ is
written as 
\begin{equation*}
\mathcal{L}_{eg}^{\prime }=-i\omega \left( \mathcal{M}+\mathcal{P}+1\right)
+k\left( 2\mathcal{J-M-P}\right) -i\left( F\mathcal{X}_{-}-F^{\ast }\mathcal{%
Y}_{-}\right) =\mathcal{L}_{eg}+\mathcal{L}_{S},
\end{equation*}
where $\mathcal{L}_{eg}$ is given in Eq. (\ref{liouvillians}).

Let us consider the algebra generated by the following set of superoperators 
$\left\{ \mathcal{L}_{eg},F\mathcal{X}_{-}-F^{\ast }\mathcal{Y}_{-},F%
\mathcal{X}_{+}-F^{\ast }\mathcal{Y}_{+},1\right\} $. The non-null
commutation relations between these elements are 
\begin{align}
\left[ \mathcal{L}_{eg},F\mathcal{X}_{-}-F^{\ast }\mathcal{Y}_{-}\right]
&= -i\omega \left( F\mathcal{X}_{+}-F^{\ast }\mathcal{Y}_{+}\right) -k\left(
F\mathcal{X}_{-}-F^{\ast }\mathcal{Y}_{-}\right) ,  \label{comm4} \\
\left[ \mathcal{L}_{eg},F\mathcal{X}_{+}-F^{\ast }\mathcal{Y}_{+}\right]
&= -\left( 2k+i\omega \right) \left( F\mathcal{X}_{-}-F^{\ast }\mathcal{Y}%
_{-}\right)  \notag \\
& +k\left( F\mathcal{X}_{+}-F^{\ast }\mathcal{Y}_{+}\right) ,  \notag \\
\left[ F\mathcal{X}_{-}-F^{\ast }\mathcal{Y}_{-},F\mathcal{X}_{+}-F^{\ast }%
\mathcal{Y}_{+}\right] &= 4\left| F\right| ^{2}.  \notag
\end{align}
The Lie exponential $\exp \left( \mathcal{L}_{eg}^{\prime }t\right) $ can be
expressed as 
\begin{align}
\exp \left( \mathcal{L}_{eg}^{\prime }t\right) &= e^{z\left( t\right)
}\exp \left[ p\left( t\right) \left( F\mathcal{X}_{-}-F^{\ast }\mathcal{Y}%
_{-}\right) \right] \exp \left[ q\left( t\right) \left( F\mathcal{X}%
_{+}-F^{\ast }\mathcal{Y}_{+}\right) \right]  \label{Lie_exp4} \\
& \times \exp \left[ s\left( t\right) \mathcal{L}_{eg}\right] ,  \notag
\end{align}
where the functions to be determined $z$, $p$, $q$ and $s$ obey the initial
condition 
\begin{equation}
z\left( 0\right) =p\left( 0\right) =q\left( 0\right) =s\left( 0\right) =0.
\label{initial_cond2}
\end{equation}
The differentiation of (\ref{Lie_exp4}) with respect to $t$ and the
successive application of Baker-Hausdorff formula (\ref{BH_formula1},\ref
{BH_formula2}) yield 
\begin{align*}
\mathcal{L}_{eg}^{\prime } &= \overset{\cdot }{z}+4\overset{\cdot }{q}%
p\left| F\right| ^{2}-2\overset{\cdot }{s}q^{2}\left( 2k+i\omega \right)
\left| F\right| ^{2}+2i\overset{\cdot }{s}\omega p^{2}\left| F\right| ^{2}-4%
\overset{\cdot }{s}kpq\left| F\right| ^{2} \\
& +\left[ \overset{\cdot }{p}+\overset{\cdot }{s}q\left( 2k+i\omega \right) +%
\overset{\cdot }{s}kp\right] \left( F\mathcal{X}_{-}-F^{\ast }\mathcal{Y}%
_{-}\right) \\
& +\left[ \overset{\cdot }{q}+i\overset{\cdot }{s}\omega p-\overset{\cdot }{s%
}kq\right] \left( F\mathcal{X}_{+}-F^{\ast }\mathcal{Y}_{+}\right) +\overset{%
\cdot }{s}\mathcal{L}_{eg}.
\end{align*}
This identity yields the following set of differential equations 
\begin{gather}
\overset{\cdot }{s} = 1,  \label{diff_eq2} \\
\overset{\cdot }{q}-\overset{\cdot }{s}\left( kq-i\omega p\right) = 0, 
\notag \\
\overset{\cdot }{p}+\overset{\cdot }{s}\left[ q\left( 2k+i\omega \right) +kp%
\right] = -i,  \notag \\
\overset{\cdot }{z}+4\overset{\cdot }{q}p\left| F\right| ^{2}+2\overset{%
\cdot }{s}\left| F\right| ^{2}\left[ i\omega p^{2}-q^{2}\left( 2k+i\omega
\right) -2kpq\right] = 0.  \notag
\end{gather}
Taking the initial condition (\ref{initial_cond2}) into account, the
solution of the above set of differential equations is 
\begin{gather}
s\left( t\right) = t,  \label{sol_diff_eq2} \\
q\left( t\right) = -\frac{\omega }{\left( k+i\omega \right) ^{2}}\left\{
\cosh \left[ \left( k+i\omega \right) t\right] -1\right\} ,  \notag \\
p\left( t\right) = i\frac{k}{\left( k+i\omega \right) ^{2}}\left\{ \cosh %
\left[ \left( k+i\omega \right) t\right] -1\right\} -i\frac{\sinh \left(
k+i\omega \right) t}{k+i\omega },  \notag \\
\begin{split}
z\left( t\right) &=-\frac{2i\omega \left| F\right| ^{2}}{\left( k+i\omega
\right) ^{2}}\left\{ t+\frac{4\left[ e^{-\left( k+i\omega \right) t}-1\right]
-e^{-2\left( k+i\omega \right) t}+1}{2\left( k+i\omega \right) }\right. 
\notag \\
& \left. +i\frac{\omega }{\left( k+i\omega \right) ^{2}}\left\{ \cosh \left[
\left( k+i\omega \right) t\right] -1\right\} ^{2}\right\} .  \notag
\end{split}
\end{gather}

Since $\mathcal{X}_{-}$ and $\mathcal{Y}_{-}$ commutate (as well as $%
\mathcal{X}_{+}$ and $\mathcal{Y}_{+}$) we rewrite (\ref{Lie_exp4}) as 
\begin{align*}
\exp \left( \mathcal{L}_{eg}^{\prime }t\right) &= e^{z\left( t\right)
}\exp \left[ Fp\left( t\right) \mathcal{X}_{-}\right] \exp \left[ -F^{\ast
}p\left( t\right) \mathcal{Y}_{-}\right] \exp \left[ Fq\left( t\right) 
\mathcal{X}_{+}\right] \\
& \times \exp \left[ -F^{\ast }q\left( t\right) \mathcal{Y}_{+}\right] \exp %
\left[ \mathcal{L}_{eg}t\right] .
\end{align*}
The commutation of the exponentials which contain the superoperators $%
\mathcal{Y}_{-}$ and $\mathcal{X}_{+}$ in the above expression yields 
\begin{align*}
\exp \left( \mathcal{L}_{eg}^{\prime }t\right) &= \exp \left[ z\left(
t\right) +2\left| F\right| ^{2}p\left( t\right) q\left( t\right) \right]
\exp \left[ Fp\left( t\right) \mathcal{X}_{-}\right] \exp \left[ Fq\left(
t\right) \mathcal{X}_{+}\right] \\
&\times \exp \left[ -F^{\ast }p\left( t\right) \mathcal{Y}_{-}\right] \exp %
\left[ -F^{\ast }q\left( t\right) \mathcal{Y}_{+}\right] \exp \left[ 
\mathcal{L}_{eg}t\right] .
\end{align*}
Hence, the expression for $\hat{\rho }_{eg}\left( t\right) $ can be put
in the form 
\begin{align*}
\hat{\rho }_{eg}\left( t\right) &= \exp \left[ z\left( t\right) +2\left|
F\right| ^{2}p\left( t\right) q\left( t\right) \right] \exp \left[ F\left(
p+q\right) \left( t\right) \hat{a}^{\dagger }\right] \exp \left[ F^{\ast
}\left( p-q\right) \left( t\right) \hat{a}\right] \\
& \times \left[ \exp \left( \mathcal{L}_{eg}t\right) \hat{\rho }%
_{eg}\left( 0\right) \right] \exp \left[ -F^{\ast }\left( p+q\right) \left(
t\right) \hat{a}\right] \exp \left[ -F\left( p-q\right) \left( t\right) 
\hat{a}^{\dagger }\right] .
\end{align*}
After some algebra, we found 
\begin{align}
\hat{\rho }_{eg}\left( t\right) &= \exp \left[ z\left( t\right) +\left|
F\right| ^{2}\left( p^{2}-q^{2}+2pq+\left| p+q\right| ^{2}\right) \left(
t\right) \right] \hat{D}\left[ \beta _{e}\left( t\right) \right]  \notag
\\
& \times \exp \left\{ 2F^{\ast }\left[ \mathop{\rm Re}\nolimits p\left(
t\right) -i\mathop{\rm Im}\nolimits q\left( t\right) \right] \hat{a}%
\right\} \left[ \exp \left( \mathcal{L}_{eg}t\right) \hat{\rho }%
_{eg}\left( 0\right) \right]  \label{rhoeg3} \\
& \times \exp \left\{ -2F\left[ \mathop{\rm Re}\nolimits p\left( t\right) -i%
\mathop{\rm Im}\nolimits q\left( t\right) \right] \hat{a}^{\dagger
}\right\} \hat{D}^{\dagger }\left[ \beta _{g}\left( t\right) \right] , 
\notag
\end{align}
where $\beta _{e}\left( t\right) $ and $\beta _{g}\left( t\right) $ are the
amplitudes defined in (\ref{betae}), (\ref{betag}). In Ref. \cite{faria},
the action of the exponential $\exp \left( \mathcal{L}_{eg}t\right) $ on an
initial condition proportional to a coherent state has been evaluated. We
employed the results obtained there to find the time development of $%
\hat{\rho }_{eg}$ corresponding to the initial state (\ref{initial_state}%
) studied here.

\acknowledgments{We gratefully acknowledge fruitfull discussions with
M. Fran{\cc}a Santos, P. Nussenzveig, M. O. Terra Cunha, M. C. de Oliveira,
M. H. Y. Moussa and S. S. Mizrahi. This work was partially supported
by the brazilian agencies CNPq (MCN) and FAPESP (JGPF).}

\newpage

{\large \textbf{Figures}}

\medskip

\begin{figure}[h]
\includegraphics[scale=.5,angle=-90]{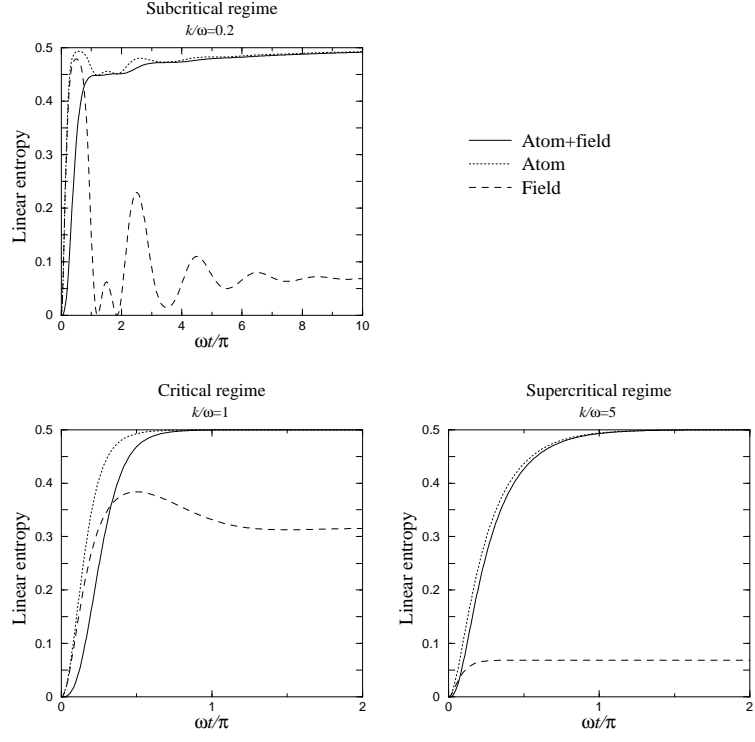}
\caption{Linear entropy of the systems atom-field (solid line),
atom (dotted line) and field (dashed line) as a function of $\omega t$ for
different values of the ratio $k/\omega $. For all plots, we have $\left|
F\right| /k=1$.}
\label{fig1}
\end{figure}

\begin{figure}
\includegraphics[scale=.5,angle=-90]{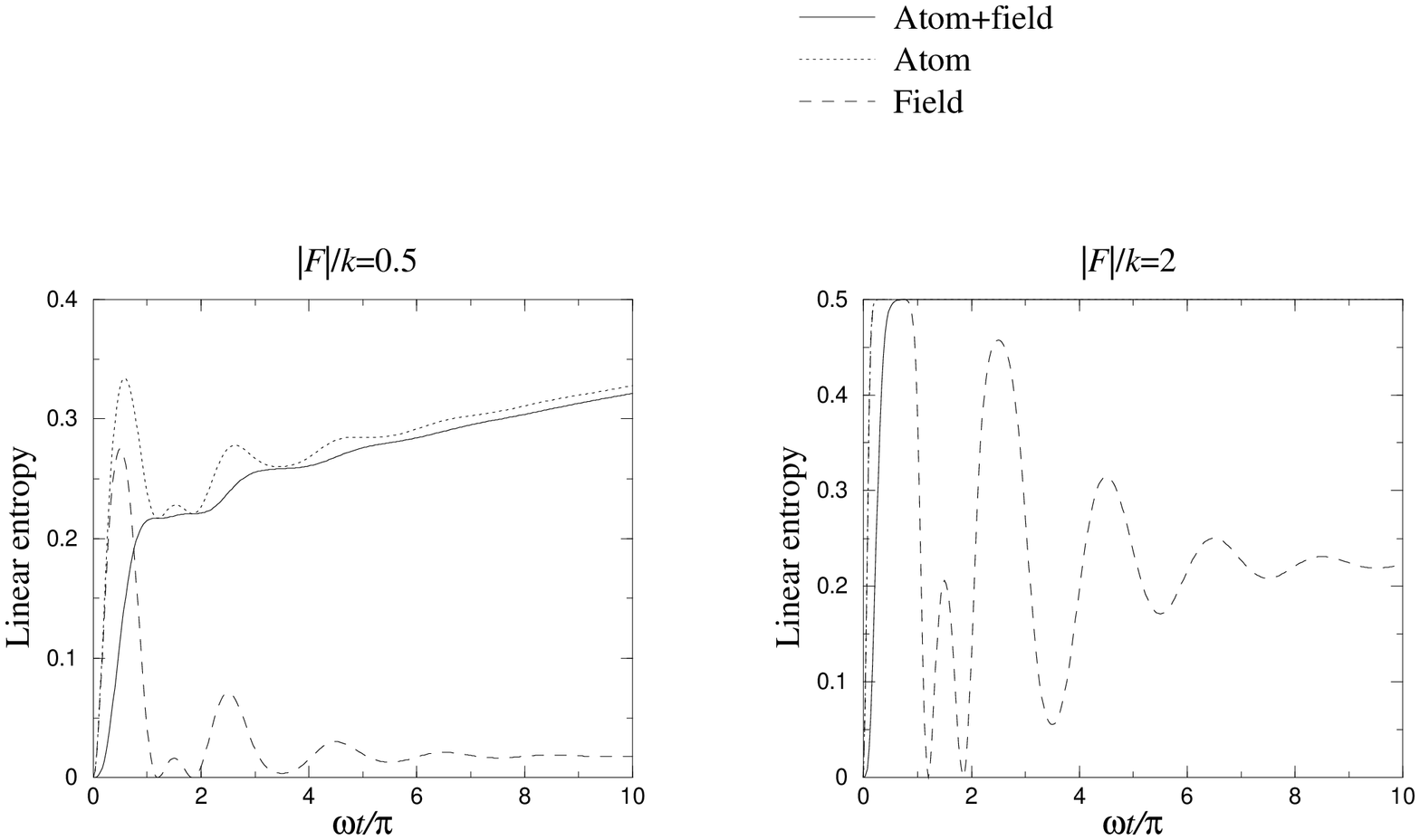}
\caption{Linear entropy of the systems atom-field (solid line),
atom (dotted line) and field (dashed line) as a function of $\omega t$ for
different values of the ratio $\left| F\right| /k$. For all plots, we have $%
k/\omega =0.2$.}
\label{fig2}
\end{figure}

\begin{figure}
\includegraphics[scale=.5,angle=-90]{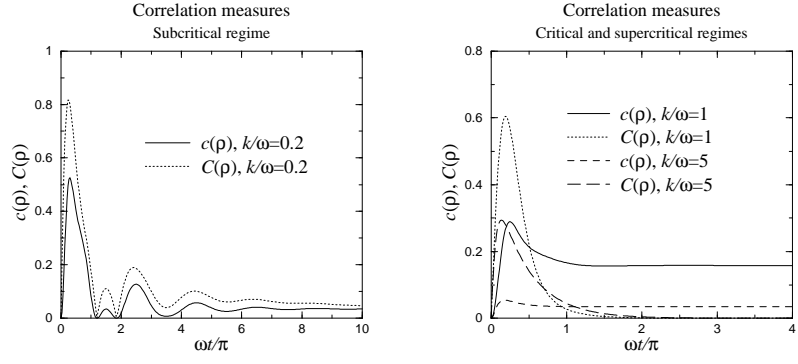}
\caption{Total correlation measure $c$ and concurrence $C$ of the
global state $\hat{\rho }$ as a function of $\omega t$. For all plots, we
have $\left| F\right| /k=1$.}
\end{figure}

\end{document}